\documentclass[9pt,twocolumn,twoside]{pnas-new}

\templatetype{pnasresearcharticle} 

\usepackage{bm}
\usepackage{amsmath}
\usepackage[capitalise]{cleveref}
\usepackage{float}
\usepackage[version=3]{mhchem} 

\usepackage{subcaption}
\newsavebox{\tallerimage}

\graphicspath{
	{../include-draft/}
}

\usepackage[single]{acro} 

\newcommand{\ea}{\textit{et al.}}
\newcommand{\ie}{\textit{i.e.}}
\newcommand{\eg}{\textit{e.g.}}

\newcommand{\bs}[2]{\mbox{#1\hspace{0.1em}\raisebox{0.15ex}{--}\hspace{0.1em}}#2}
\newcommand{\bd}[2]{\mbox{#1\hspace{0.3em}\raisebox{0.18ex}{=\hspace{-0.8em =}}\hspace{0.3em}}#2}
\newcommand{\bw}[2]{\mbox{#1\hspace{0.02em}\raisebox{0.6ex}{.\hspace{-0.07em}.\hspace{-.07em}.}\hspace{0.03em}}#2}

\newcommand{\TWcites}[1][]{{\color{red} \textbf{[\emph{\ifthenelse{\isempty{#1}}{cites}{\emph{cite:} #1}}]}}}

\title{Observing the 3D chemical bond and its energy distribution in a projected space}

\author[a]{Timothy R.~Wilson}
\author[a]{Malavikha Rajivmoorthy}
\author[a]{Jordan Goss}
\author[a]{Sam Riddle}
\author[a,1]{M.~E.~Eberhart}

\affil[a]{Molecular Theory Group, Colorado School of Mines, 1500 Illinois St., Golden, Colorado, USA}

\leadauthor{Wilson}

\authordeclaration{The authors have no conflicts of interest to declare.}
\correspondingauthor{\textsuperscript{1}To whom correspondence should be addressed. E-mail: \href{mailto:meberhar@mines.edu?subject=Observing\%20the\%203D\%20chemical\%20bond\%20and\%20its\%20energy\%20distribution\%20in\%20a\%20projected\%20space\%20(Arxiv.org)}{meberhar@mines.edu}}

\keywords{\textit{Keywords:} bond bundle $|$ gradient bundle $|$ condensed charge density $|$ bond theory $|$ computational chemistry} 
	
\begin{abstract}
	Our curiosity-driven desire to ``see'' chemical bonds dates back at least one-hundred years, perhaps to antiquity.
	Sweeping improvements in the accuracy of measured and predicted electron charge densities, alongside our largely bondcentric understanding of molecules and materials, heighten this desire with means and significance.
	Here we present a method for analyzing chemical bonds and their energy distributions in a two-dimensional projected space called the condensed charge density.
	Bond ``silhouettes'' in the condensed charge density can be reverse-projected to reveal precise three-dimensional bonding regions we call bond bundles.
	We show that delocalized metallic bonds and organic covalent bonds alike can be objectively analyzed, the formation of bonds observed, and that the crystallographic structure of simple metals can be rationalized in terms of bond bundle structure.
	Our method also reproduces the expected results of organic chemistry, enabling the recontextualization of existing bond models from a charge density perspective.
\end{abstract}

\dates{This manuscript was compiled on \today}
\doi{DOI: ChemPhysChem \href{https://doi.org/10.1002/cphc.201900962}{cphc.201900962}}

\begin{document}

\maketitle
\thispagestyle{firststyle}
\acresetall
\acuse{ncp,bcp,rcp,ccp}
\ifthenelse{\boolean{shortarticle}}{\ifthenelse{\boolean{singlecolumn}}{\abscontentformatted}{\abscontent}}{}

\hrule
\tableofcontents\vspace{1.2em}
\hrule

\section{Introduction}

\dropcap{T}he great statistician George Box reputedly remarked, ``All models are wrong but some are useful'' \cite{Box_wrong_models}.
Box's observation has been referenced as relevant to scientific models in general and is particularly germane to the chemical sciences \cite{Box1979, Hoffmann_on_comp_chem, Eberhart_foundations1, Eberhart_foundations2, Anti_qtaim2006b}.
After all, much of chemistry relies on empirical models that have survived by proving useful to those creating new molecules and materials.
Foremost among these are representations of chemical bonding that now undergird all of chemistry.

Bonding models are useful when providing a framework from which to estimate energy differences due to subtle changes in the arrangement or composition of an atomic system.
Such useful models have a venerable history, arguably beginning with Gilbert Lewis' century old insights regarding electron sharing \cite{lewis1916}, which later formed the kernel of the valence bond theory of Slater and Pauling \cite{Slater_Pauling_VBT}.
The evolving perspectives of chemical bonding are almost too numerous to mention and constitute a significant portion of the chemical literature.
However, the contributions to this corpus by such luminaries as Mulliken, Hammond, Coulson, Fukui, Hoffmann, Ruedenberg, H\"{u}ckel, Goddard, Pople, Parr, Peyerimhoff, Karplus, Levitt, and Warshel \cite{mulliken1967,hammond1955,coulson1949,fukui1982,hoffmann1982,Ruedenberg1962,hueckel1931,tannor1994,pople1999,parr1989,buenker1974,mccammon1977,levitt1975,warshel1997} demand recognition.

Despite the advantages current bonding models confer, they have proven difficult to apply broadly, that is, to all the stuff held together by ``bonds.''
For example, to metals and alloys where suitable chemical bonding models might prove just as useful as those that have been employed in the design and synthesis of organic polymers \cite{Hoffmann_JT}.

There are forces at work that may lead to a change in this situation.
Advances in computational density functional methods \cite{DFT_review} coupled to ever more accurate measurements of the electronic charge density \cite{High_res_xray} are providing the impetus for the development of charge density based bonding models \cite{Gatti_CDA}. Because the electronic charge density is an observable---existing independently from the methods used for its calculation or measurement---such models should prove useful across all classes of molecules and solids. An observable chemical bond in particular could similarly be used to extend existing bonding models to new fields by connecting them with their roots in the charge density.

Foremost among the efforts to frame chemical principles around the charge density is Bader's \ac{qtaim} \cite{AIM, Matta_AIM}.
With its clearly constructed formalism through which to identify an atom's boundary, \ac{qtaim} brings clarity and consistency to a number of previously ill-defined chemical concepts such as the energies, sizes and electron counts of the atoms comprising a molecule or solid.
However, the topological representation of chemical bonding ensuing from this theory is plagued by an ongoing debate---both questioning \cite{Anti_qtaim2004, Anti_qtaim2006a, Anti_qtaim2006b, shahbazian2018, no_bcp_2006} and supporting \cite{matta2003, Bader_biphenyl, Matta2006_h-h, Matta2007_h-h, Pro_qtaim2008, Matta2018,pendas2007} its rigor and utility.
It should be noted, however, that the topological approach to bond analysis due to \ac{qtaim} neglects the central attribute of the theory---the partitioning of charge density into regions with well-characterized energies.
We show here that this omission is not intrinsic to \ac{qtaim}-based bond analysis approaches.

We proceed by reviewing the consequences and rationale behind charge density partitioning.
Building on this rationale, we define a maximally partitioned charge density space---the condensed charge density---in which the kinetic energy resulting from the Laplacian family of kinetic energy operators is everywhere well-defined \cite{Ayers_ambiguous_KE}.
While faithfully recovering the essential elements of \ac{qtaim}, the topology of this space reveals a charge density volume with the properties of a chemical bond, which is designated a bond bundle.
We then examine several instances in which QTAIM's topological bond has been asserted to be faulty and demonstrate that these assertions are made moot by the bond bundle construct.
We additionally argue that the condensed charge density space is ideally suited to describe chemical phenomena as its structure derives from a preferred moving coordinate frame giving primacy to charge density isosurfaces.
Supported by these findings and arguments, we apply our approach to some simple metals and demonstrate that their crystallographic structure can be rationalized as a consequence of bond bundle structure.

\section{Charge density partitioning}
 
The significance of the electronic kinetic energy as a mediator of chemical bonding was recognized as early as the 1930s, for example by Hellman \cite{Hellmann1933} and Slater \cite{Slater_virial}.
The central role of the kinetic energy took on further chemical import in Ruedenberg's classic 1962 paper, {\it The Physical Nature of the Chemical Bond} \cite{Ruedenberg1962},
which prompted subsequent efforts to capture changes to the local kinetic energy as necessary for useful theories of bonding.
These efforts were confounded, however, by the the multiple representations for local kinetic energy \cite{Bader_KE, Ayers_ambiguous_KE}.
Among the forms commonly used is one referred to here as the gradient representation, $T_G$ \cite{feinberg1971, feinberg1971a}, in which the total kinetic energy over a region $\Omega$ appears as,
\begin{equation}\label{eq:Tgrad}
	T_G(\Omega)= {\frac {\hbar^2}{4m}}N \int_\Omega d\bm{r}\int d\tau^\prime ~ \nabla \Psi^* \cdot \nabla\Psi.
\end{equation}
An alternative form for the kinetic energy of the same region may be expressed in terms of the Schr{\"o}dinger (Laplacian) kinetic energy \cite{feinberg1970} as,
\begin{equation}\label{eq:Tlap}
	T_L(\Omega)= -{\frac {\hbar^2}{4m}}N \int_\Omega d\bm{r}\int d\tau^\prime ~[\Psi \nabla^2 \Psi^* + \Psi^* \nabla^2 \Psi].
\end{equation}

In general $T_L(\Omega)$ and $T_G(\Omega)$ differ.
It is straightforward to show \cite{Bader_KE} that
\begin{equation}\label{eq:Tdiff}
	T_L(\Omega)- T_G(\Omega)= -{\frac {\hbar^2}{4m}}N \int dS(\Omega,\bm{r})\nabla \rho(\bm{r}) \cdot \bm{n}(\bm{r}), 
\end{equation}
where $\nabla \rho(\bm{r})$ is the gradient of the charge density, $S(\Omega, \bm{r})$ is the surface bounding the region $\Omega$ and $\bm{n}(\bm{r})$ is the unit vector normal to this surface at $\bm{r}$.
The integral gives the net flux of the charge density gradient through $S(\Omega, \bm{r})$.
Where this flux is zero, the two kinetic energy representations give the same value---as is required if $\Omega$ spans an entire molecule or the unit cell of a periodic solid.
Thus, the kinetic energy of a molecule or a crystal's unit cell is unambiguous.
However, smaller regions contained in a molecule or unit cell may also possess well-defined energies if the net flux of the gradient of the charge density is zero over their boundaries \cite{Shahbazian2011}, which necessarily includes regions bounded by surfaces over which the flux of $\nabla \rho(\bm{r})$ is everywhere zero.
Such regions are said to be bounded by zero flux surfaces and it was believed that each possess a well-defined kinetic energy \cite{AIM}.%
\footnote {
	Anderson \ea\ \cite{Ayers_ambiguous_KE}~have demonstrated that the kinetic energy ambiguity is much broader than embodied in \cref{eq:Tgrad,eq:Tlap}.
	We will address the broader implications of kinetic energy ambiguity in \cref{sec:content} of this paper.
}

Bader noted a unique zero flux surface surrounding the nucleus of every atom of a molecule, making it possible to associate a kinetic, potential and hence total energy ($T(\Omega)$, $V(\Omega$), $E(\Omega)$ respectively) with an atom in a molecule.
Consistent with chemical models that assume energies to be additive, the molecular energy is given as the sum of these atomic energies.
In addition to their well-characterized energies, Bader showed that these atomic regions (actually all zero flux surface bounded regions) satisfy the virial theorem \cite{AIM}.
Accordingly, for molecules or solids in structural equilibrium or at a transition state \cite{Slater_virial, Rudenberg_virial},
\begin{equation}
	E(\Omega) = \frac{1}{2} V(\Omega) = -T(\Omega).
\end{equation}
The atoms in a molecule (or solid)---called Bader atoms or atomic basins---owe their energetic significance to the zero flux surfaces that bound and define them.
Obviously, The \acl{qtaim} derives its name from this fundamental association.

Explicit atomic boundaries impose a connectivity on the atoms of a molecule.
Specifically, two atomic basins that touch over some finite region of their mutual boundary may be connected by an observable ridge of maximum charge density that extends from one atomic nucleus to the other.
Bader called this ridge a ``bond path.''
A necessary condition for the existence of such a path is a charge density saddle point of index $-1$ on their shared boundary, which Bader named a \ac{bcp}.

For the molecules Bader originally investigated, their sets of bond paths recovered Lewis' molecular graphs.
However, subsequent and ongoing investigations (see for example References \citenum{Anti_qtaim2006a} and \citenum{matta2003}) have recovered bond paths and \acp{bcp} between atoms that the Lewis model does not predict to be ``bound.'' 
In some instances bond paths have been discovered between atoms that are argued to be repulsive \cite{Pro_qtaim2008}.
And in still other instances bond paths are absent between atoms that chemical reasoning predicts to be bound \cite{no_bcp_2006, no_bcp_2013a, no_bcp_2013b, no_bcp_2007, no_bcp_2009}.
As mentioned, such revelations have prompted questions regarding the extent to which a bond path should be taken as an indicator of a bond and the indicated interaction as ``bonding" (see particularly References \citenum{Anti_qtaim2004, Anti_qtaim2006a} and \citenum{shahbazian2018}).

Pend\'{a}s \ea\ used a wave function-based interacting quantum atoms approach to investigate the selection mechanism for bond paths.
They showed that bond paths indicate the channels along which quantum exchange energy is most lowered \cite{pendas2007}.
A bond path may still correspond to stable, unstable, or metastable interactions depending on the role played by exchange, and their work offers a clue as to why bond paths may be found in some instances but not others.

The larger debate, however, assumes that connections between the atoms of a molecule must be stabilizing with respect to total energy.
For the vast majority of organic molecules, which conform to the Lewis model, stabilizing effects proceed through electron sharing and hence connectivity.
However, we see no {\it a priori} reason that this must be the case and believe that structures giving rise to stability must be embedded in a space in which energy is everywhere well defined.
This is the space of all charge density volumes bound by zero flux surfaces called the \ac{P}.

\subsection{Condensed charge density space} 
\label{sec:condencedRhoSpace}

\begin{figure}
	\centering
	\includegraphics[width=0.6\linewidth]{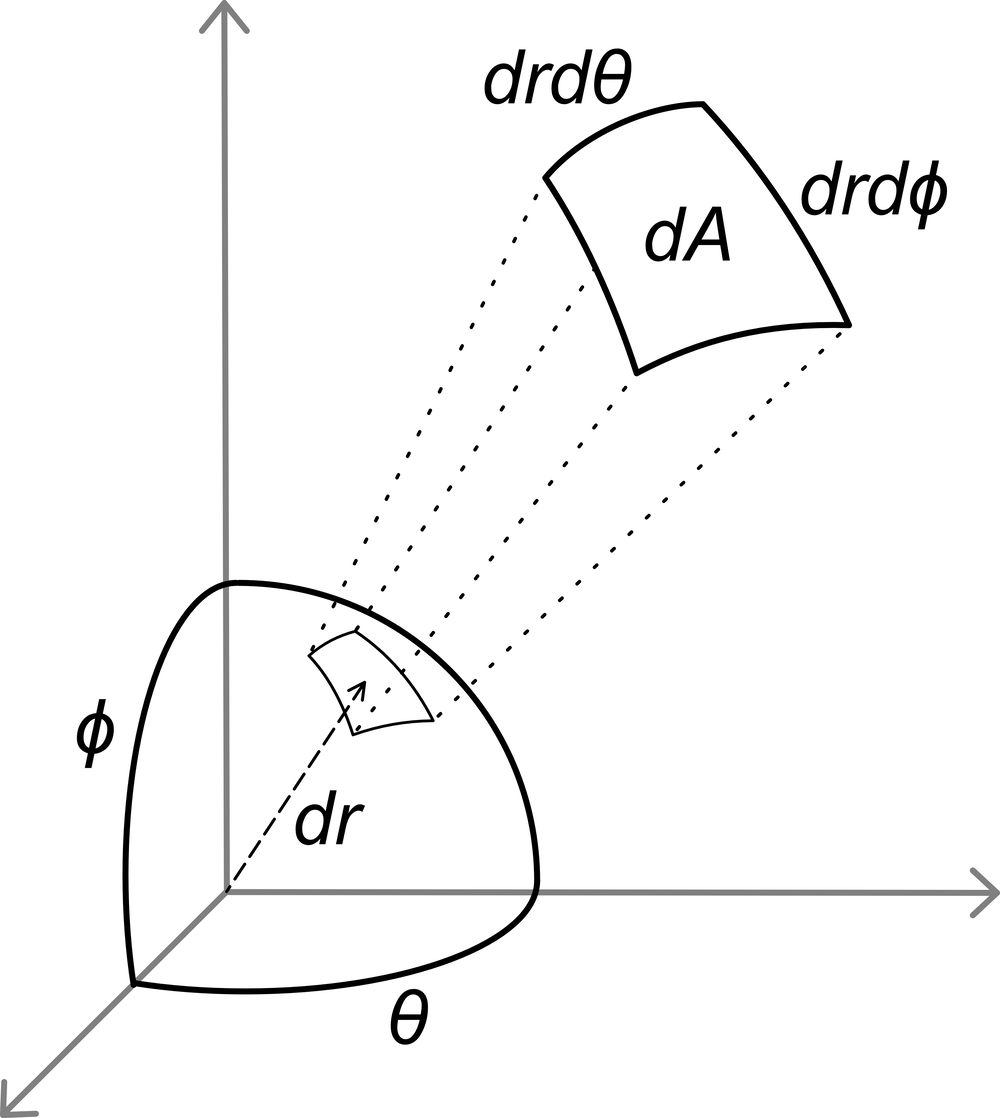}
	\caption{\label{fig:dA}%
		Differential area element on a sphere.
	}
\end{figure}

\begin{figure}
	\centering
	\includegraphics[width=0.9\linewidth]{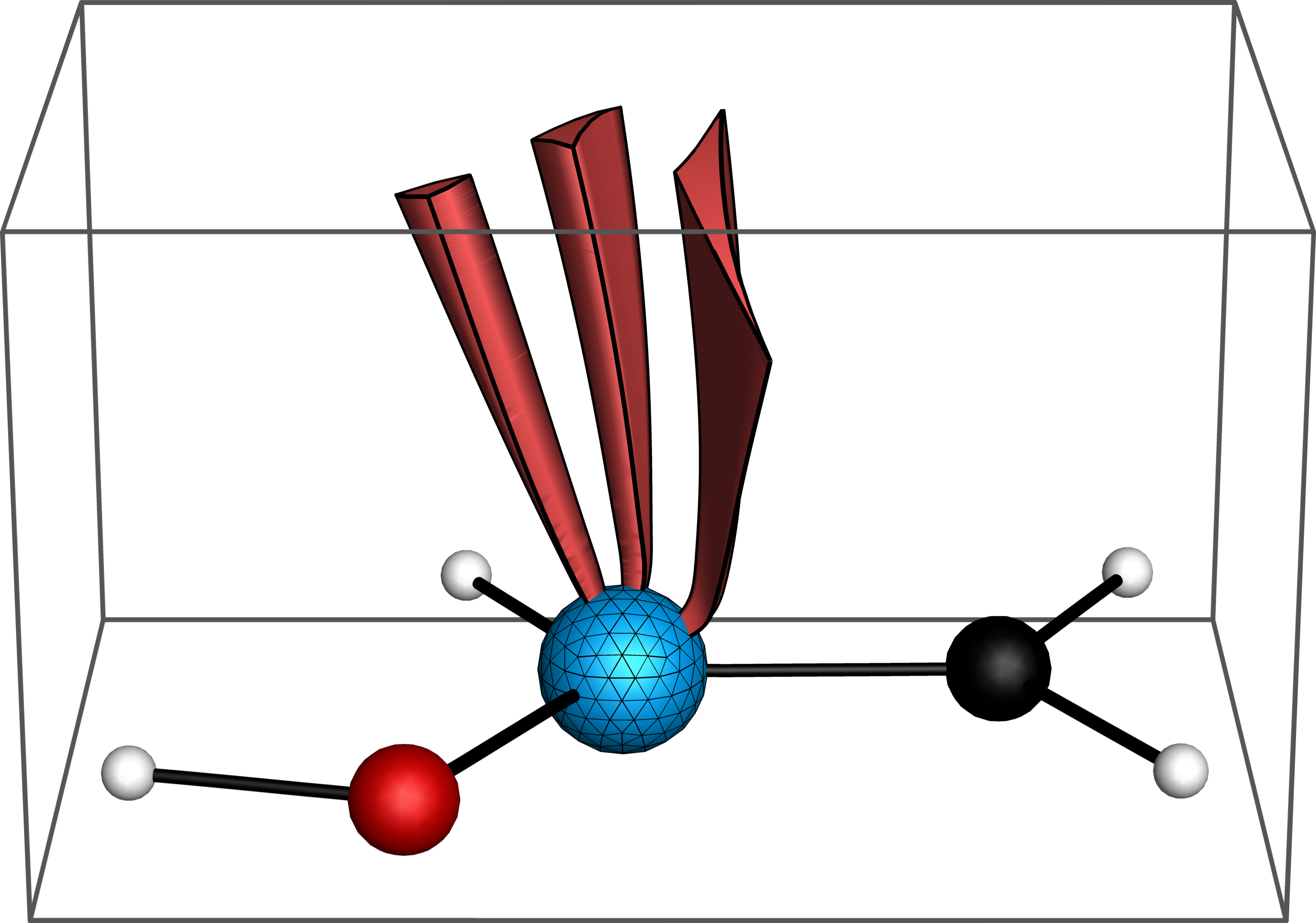}
	\caption{\label{fig:dgb}%
		A sampling of differential gradient bundles.
	}
\end{figure}

\begin{figure*}
	\centering
	\includegraphics[width=0.9\linewidth]{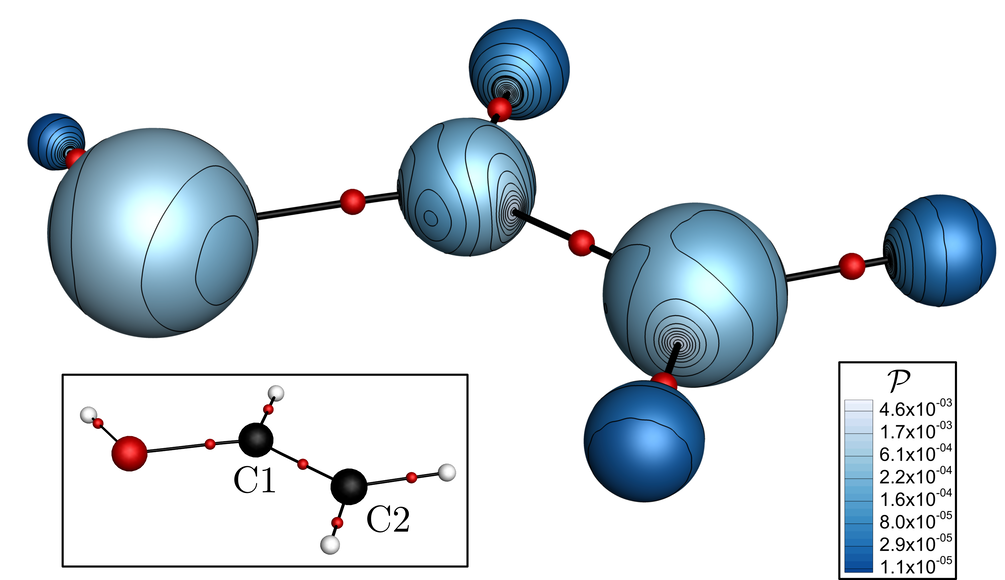}
	\caption{\label{fig:gba-mol-atlas}%
		\acs{P} for all atoms in vinyl alcohol (\ie\ its molecular atlas).
		Inset: Black, white, and red spheres respectively indicate carbon, hydrogen, and oxygen nuclear positions, with black paths and small red spheres denoting bond paths and \acp{bcp} (same scheme used when appropriate in remaining figures).
	}
\end{figure*}

\begin{figure}
	\centering
	\includegraphics[width=\linewidth]{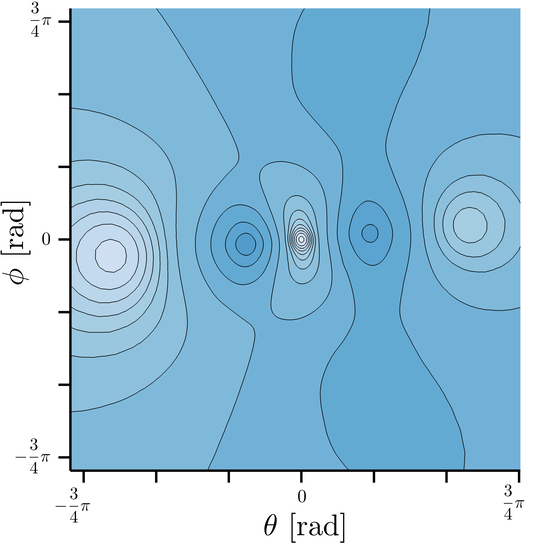}
	\caption{\label{fig:gba-projection-square}%
		Stereographic projection with contours of \ac{P} for the C1 in vinyl alcohol atom with the \bd{C}{C} bond path at the origin.
		Axes are in units of radians, corresponding to rotation around the sphere in \cref{fig:gba-mol-atlas}.
		For the remaining figures, contour shading is such that values increase from blue to white.
	}
\end{figure}

As we summarized in an earlier article \cite{wilson2019}, the space \ac{P} is constructed through a mapping of \aclp{gp} \acreset{gp} of \ac{rho} to points in \ac{P}.
Every \ac{gp} in \ac{rho} originates from a local minimum called a \ac{ccp} and terminates at a maximum, almost always coincident with a nucleus and hence denoted a \ac{ncp}.
We assume these \acp{gp} to be parameterized by arc length, $s$.

Sufficiently close to their terminus \acp{gp} are radial, making it convenient to imagine every \ac{ncp} as the center of a sphere S$_i$ of radius $dr$ (in practice $dr \approx 0.2$\AA).
Passing through every point on the surface of these spheres is a \ac{gp}.
The points on such a sphere may be specified by a polar and an azimuthal angle, so each of the molecule's \acp{gp} may be specified by a pair of coordinates and the index of the \ac{ncp} at its terminus, \ie\ \ac{gp}$_i(\theta,\phi)$.
In this reference system, the charge density of the atomic basin $i$ is a function of $ \theta, \phi$ and $s$, 
where $\theta$ and $\phi$ pick out a unique gradient path in the atomic basin $i$, and $s$ is the euclidean distance along this path.

Imagine covering each S$_i$ with a set of nonintersecting differential elements of area $dA = dr^2 \, d\theta \, d\phi$ (\cref{fig:dA}).
The \acp{gp} passing through the points interior to each of these area elements give rise to a family of infinitesimal volume elements called \acuse{dgb}\aclp{dgb}, \aca{dgb} \cite{GBA1,GBA2}, each of which is bounded by a zero flux surface.
Significantly (see \cref{sec:content}), the cross sectional areas of these \aca{dgb} (\cref{fig:dgb}) change throughout their length and thus $dA$ is a function of $s$, $\theta$, and $\phi$. 
The union of all \ac{dgb}$_i$ is equivalent to the union of all \acp{gp} terminating at \ac{ncp} $i$ and hence recovers Bader's atomic basin. And significantly, these are the smallest structures bounded by zero flux surfaces and accordingly possessing well-defined energies and energy related properties.


Explicitly, for any scalar field, $f_i$, there exists a corresponding condensed scalar field, ${\cal F}_i$, that is a function of $\theta$ and $\phi$ and a functional of $f_i$, such that
\begin{equation}
\label{eq:condensed_property}
{\cal F}[f_i]\equiv{\cal F}_i[\theta, \phi, f_i(\theta, \phi, s)]=\!\!\! \displaystyle{\int\limits_{{\rm G}_i(\theta, \phi)} \!\!\! f_i(s) dA(s) ds}.
\end{equation}
In particular, the charge density yields the condensed charge density (${\cal F}[\rho]={\cal P}$), the gradient or Laplacian kinetic energy densities yield the condensed kinetic energy density (${\cal F}[T_G]={\cal F}[T_L]={\cal T}$), and so forth.
For the special case where $f_i(\theta, \phi, s)=1$, the \acl{V} is produced (${\cal F}[1]={\cal V}$).%
\footnote{In open systems a step function---defined to be one inside some charge density isosurface (commonly taken to be $\rho = 0.001~au$) and zero outside this region---is used to obtain \ac{V}.}

As an illustrative example, \cref{fig:gba-mol-atlas} depicts \ac{P} for each of the $i$ atoms of vinyl alcohol, where every point on the sphere surrounding an atom is colored to represent the magnitude of the integrated electron density in the \ac{dgb} originating at that point.
Borrowing terminology from differential geometry, each \ac{P}$_i$ is referred to as an \textit{atomic chart} and the set of all atomic charts comprising a molecule is termed its \textit{molecular atlas}.

As an alternative to representing atomic charts on spheres, they may be projected onto a flat space as in \cref{fig:gba-projection-square} where the alcohol carbon (C1) atomic chart is represented with a stereographic projection centered at the \bd{C}{C} bond path.

By construction, every point in \ac{P} maps to a \ac{gp} in \ac{rho}.
Hence every trajectory through \ac{P} maps to a zero flux surface in \ac{rho}, and any closed loop in \ac{P} maps to a volume in \ac{rho} bounded by a zero flux surface and thus characterized by a well-defined energy.
Such volumes are called gradient bundles \cite{GBA1, GBA2, Jordan1}.
All previously noted zero flux surface-bounded volumes, \eg\ atomic basins, are proper subsets of the space of gradient bundles.


\section{Bonds in molecules}

\subsection{The topology of the condensed charge density}
\label{sec:condencedRhoTopology}

Maxima in \ac{P} typically map to bond paths in \ac{rho}, as is the case for vinyl alcohol and evident in \cref{fig:gba-mol-atlas} where each of the molecule's bond paths (shown as black lines) intersects an atomic chart at a maximum in \ac{P}.

Just as all \acp{gp} terminating at the same maximum in \ac{rho} delineate an atomic basin as a unique volume, \acf{Pgp} 
terminating at the same maximum define a unique two-dimensional basin in \ac{P} and hence a unique zero flux surface-bounded volume in \ac{rho}.

\begin{figure*}[h!]
	\centering
	\savebox{\tallerimage}{\includegraphics[width=0.42\linewidth]{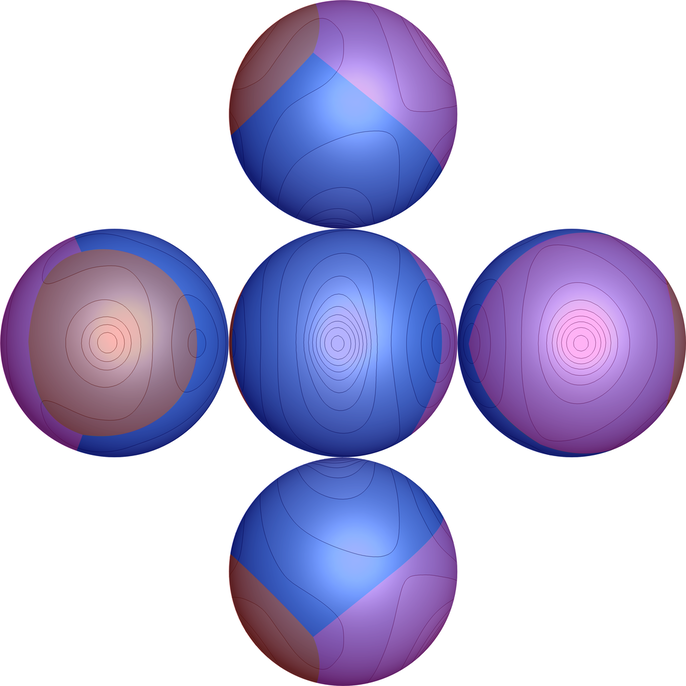}}%
	\subcaptionbox{\label{fig:gba-split-projections}}
	{\raisebox{\dimexpr.5\ht\tallerimage-.5\height}{\includegraphics[width=0.42\linewidth]{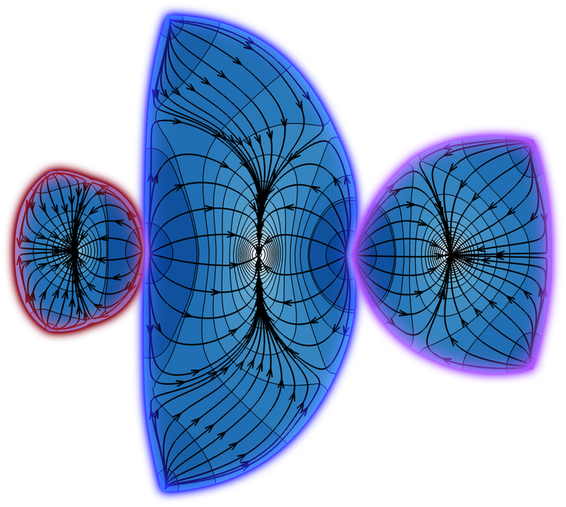}}}
	\subcaptionbox{\label{fig:gba-multi-sphere}}
	{\usebox{\tallerimage}}
	\caption{
		\textbf{\subref*{fig:gba-split-projections})} Three stereographic projections with contours of \ac{P} for the C1 atom in vinyl alcohol, centered at the \bs{C}{O} (red), \bs{C}{C} (blue), and \bs{C}{H} (magenta) bond paths and truncated at the boundaries of the respective \ac{P}-basins.
		\textbf{\subref*{fig:gba-multi-sphere})} Multiple views of a spherical mapping with contours of \ac{P} for the same C1 atom in vinyl alcohol.
		The same \bs{C}{O}, \bs{C}{C}, and \bs{C}{H} \ac{P}-basins are shaded using the same color scheme as in (a).
	}
\end{figure*}

\begin{figure*}[h!]
	\centering
	\includegraphics[width=0.88\linewidth]{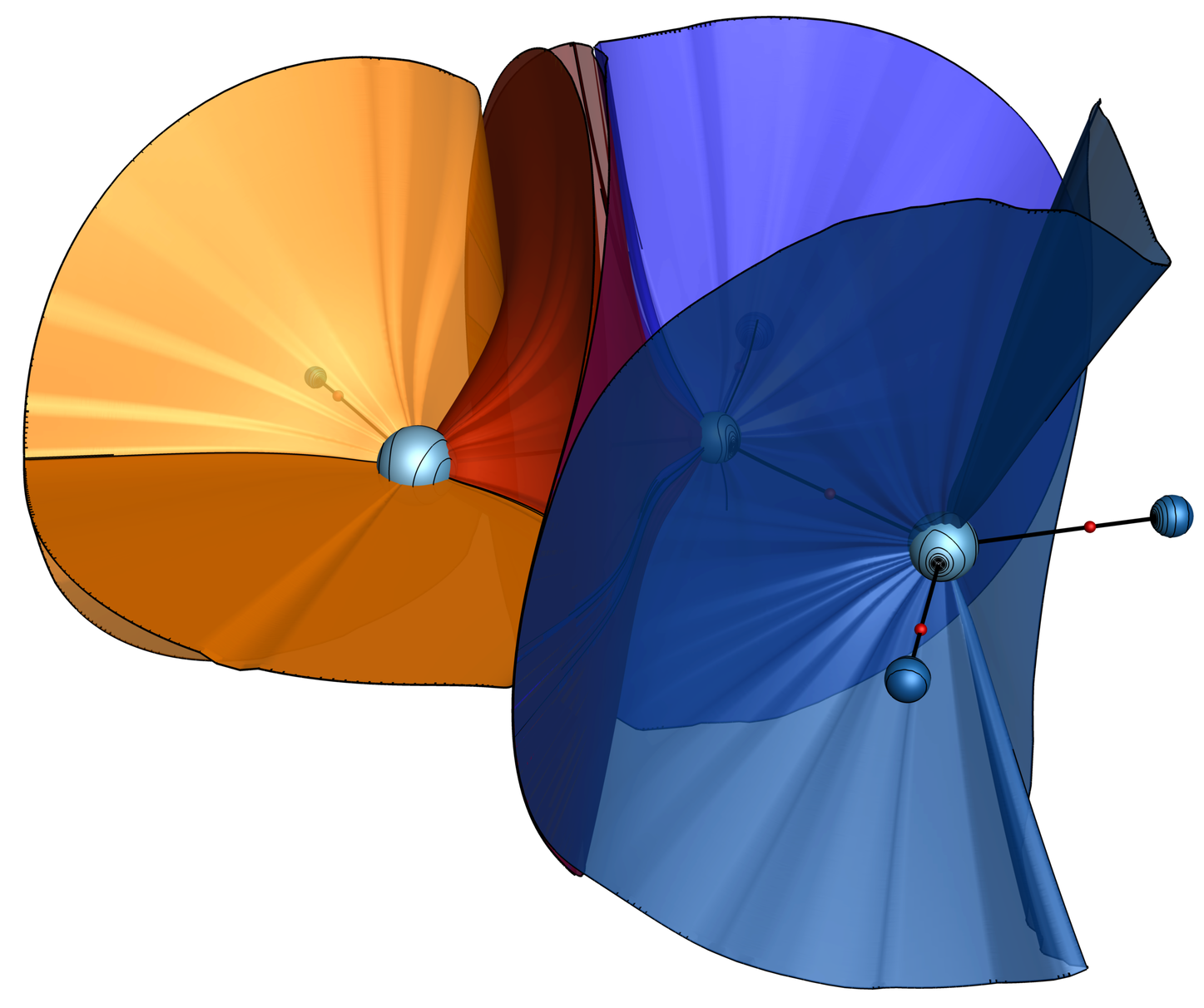}
	\caption{\label{fig:ethenolBBs}%
		\textit{Chihulyesque} bond bundle surfaces for the \bd{C}{C} (blue) and \bs{C}{O} (red) bonds and the oxygen lone pair wedge surfaces (orange) in vinyl alcohol.
	}
\end{figure*}

%
%

\begin{figure*}[h!]
	\centering
	\savebox{\tallerimage}{\includegraphics[width=.4\linewidth]{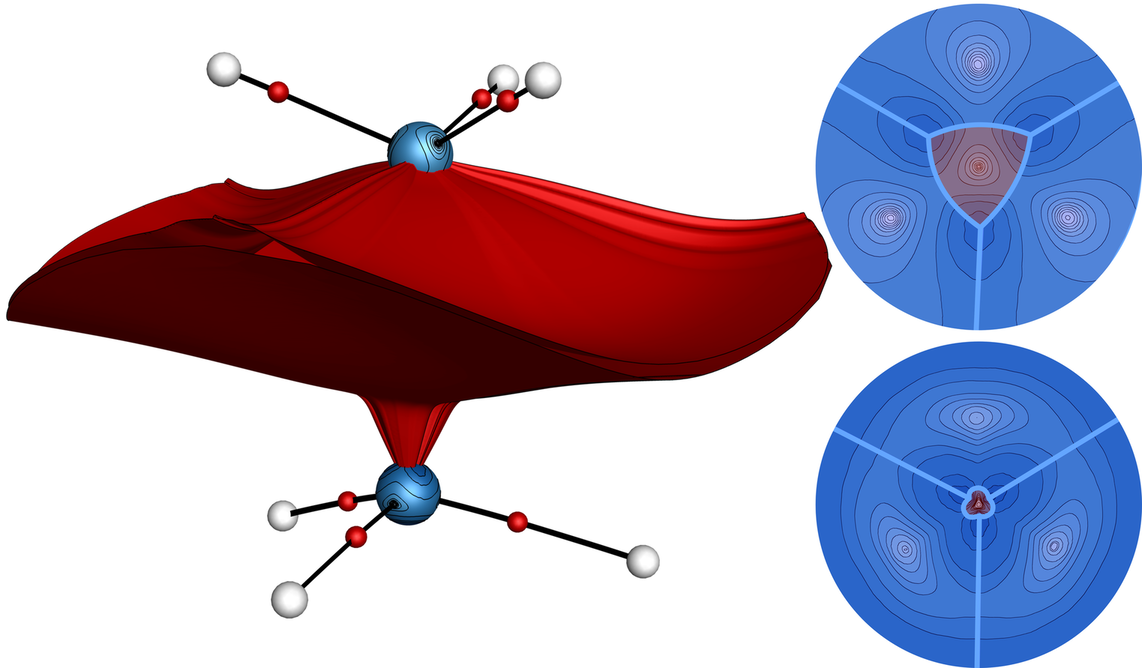}}%
	\subcaptionbox{\label{fig:BH3NH3_isolated}}
	{\raisebox{\dimexpr.5\ht\tallerimage-.5\height}{\includegraphics[width=.4\linewidth]{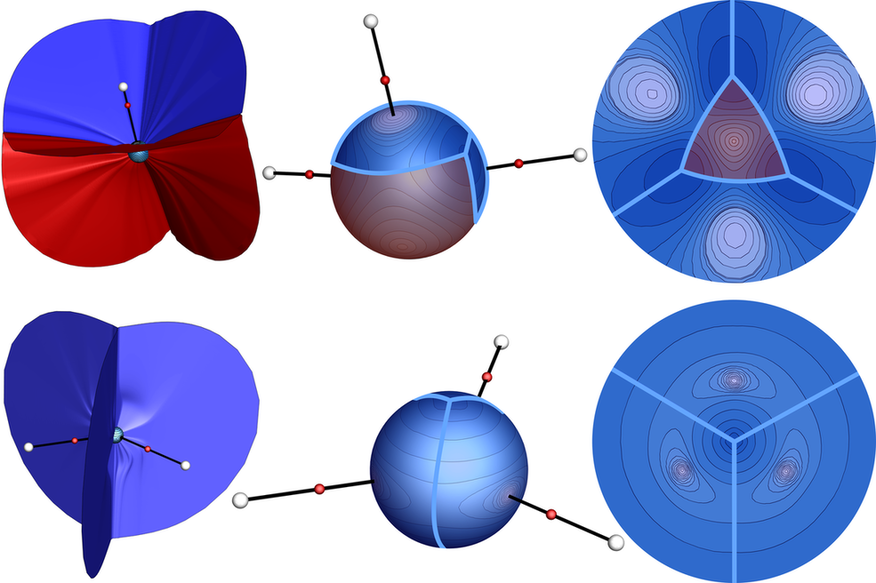}}}
	\subcaptionbox{\label{fig:H3B-NH3_product}}
	{\usebox{\tallerimage}}
	\caption{
		\ac{P}, bond bundle, and bond wedges for ammonia and borane before and after reacting.
		\textbf{\subref*{fig:BH3NH3_isolated})} \ac{P} for nitrogen in ammonia (top) and for boron in borane (bottom).
		Left column shows the 3D bond bundle (and wedge) surfaces.
		The center and right columns show spherical and projected mappings of \ac{P} with the boundaries of the bond wedges in \ac{P} delineated by blue paths.
		Projections are centered at the south and north poles of the nitrogen and boron spheres respectively.
		Nitrogen's lone pair wedge is shaded red.
		\textbf{\subref*{fig:H3B-NH3_product})} \bs{B}{N} bond bundle in \ce{H3B-NH3} with stereographic projections of \ac{P} for nitrogen (top right) centered on its south pole and boron (bottom right) centered on its north pole.
	}
\end{figure*}

\begin{figure*}[h!]
	\centering
	\includegraphics[width=.8\linewidth]{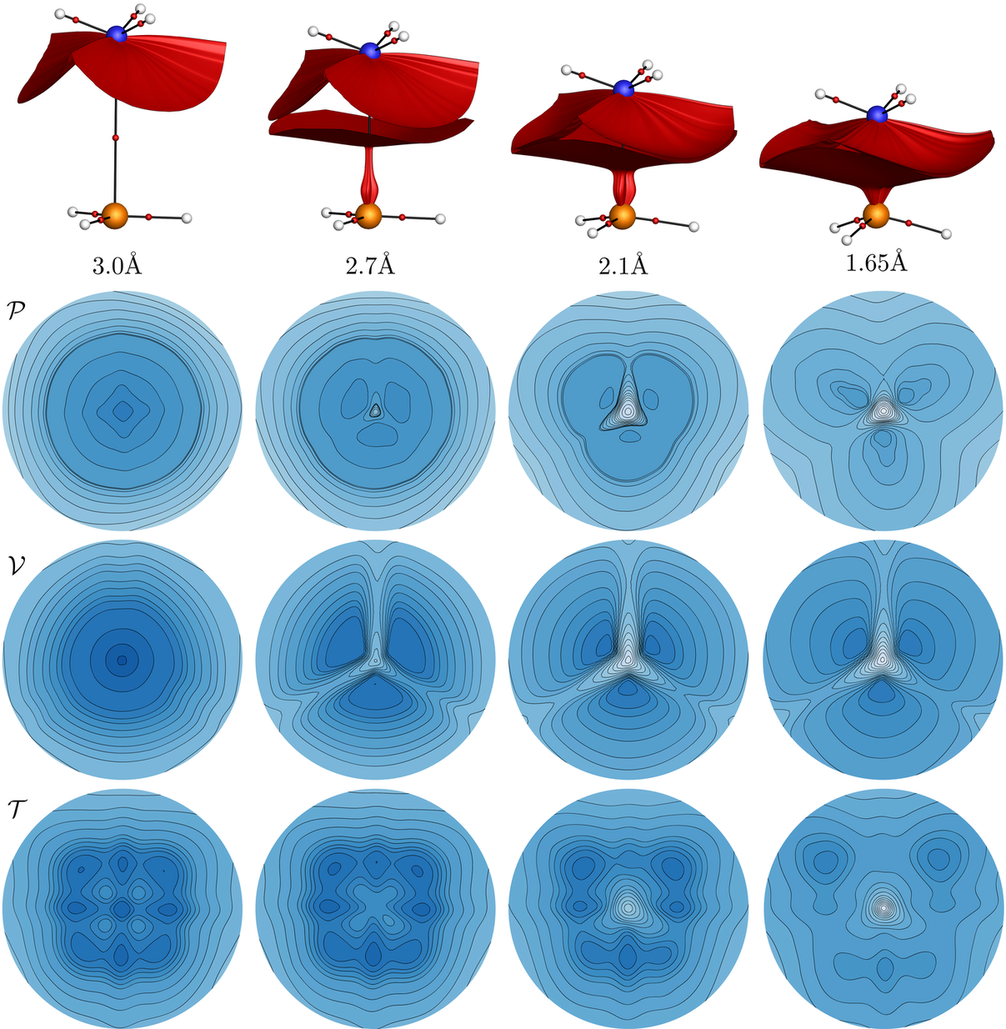}
	\caption{\label{fig:BH3NH3_b}%
		The sequence of bond bundles and corresponding projections of \ac{P}, \ac{V}, and \ac{T} centered on the north pole of the boron atom.
		The same contour values are used across each row.
		Note that the absence of three fold symmetry in the condensed properties is a computational artifact that results from the use of a rectilinear grid of \ac{rho} data.
		This artifact is particularly evident over  regions where a condensed property is nearly flat, \eg\ \ac{T} on boron in the early stages of bond bundle formation.
	}
\end{figure*}

Continuing with the vinyl alcohol example, \cref{fig:gba-split-projections} shows stereographic projections of the C1 atomic chart centered on each of its three maxima.
 Also shown are the three sets of \acsp{Pgp} terminating at each maximum and delineating corresponding \ac{P}-basins.
The image of all \ac{P}-basins of the molecular atlas partitions \ac{rho} into space-filling regions bounded by zero flux surfaces, each containing a bond path.
The energy of each region is well defined and additive to give the molecular energy.
These are the characteristics associated with a chemical bond.

Accordingly, we offer the following definitions: i) A \textit{bond wedge} is the image in \ac{rho} of \acl{Pgp} terminating at a common maximum; and ii) a \textit{bond bundle} is the union of bond wedges sharing a common intersection with an interatomic surface.%
\footnote{In the vast majority of cases the general definition provided here recovers the same regions as those resulting from an earlier bond bundle definition \cite{jones2009, jones2010}.}
In the case of lone electron pairs, the terms \textit{lone pair wedge} and \textit{lone pair bundle} can be used.
Some of the bond bundles and lone pair wedges of vinyl alcohol are shown in \cref{fig:ethenolBBs}.

Bond wedges and bond bundles are attractive in their one-to-one mapping with organic bonds as understood through valence bond theory, but they are merely the demarcation of charge density made possible through a gradient bundle decomposition of the system.
The condensed charge and energy distribution \textit{within} a bond wedge contains vastly more information, and we anticipate that these distributions will be significant in discovering the charge density structure underlying subtle differences between systems otherwise thought to operate via the same chemical mechanisms.

\subsection{Bond formation}
\label{sec:bh3nh3}

\begin{figure*}[h!]
	\centering
	\includegraphics[width=.85\linewidth]{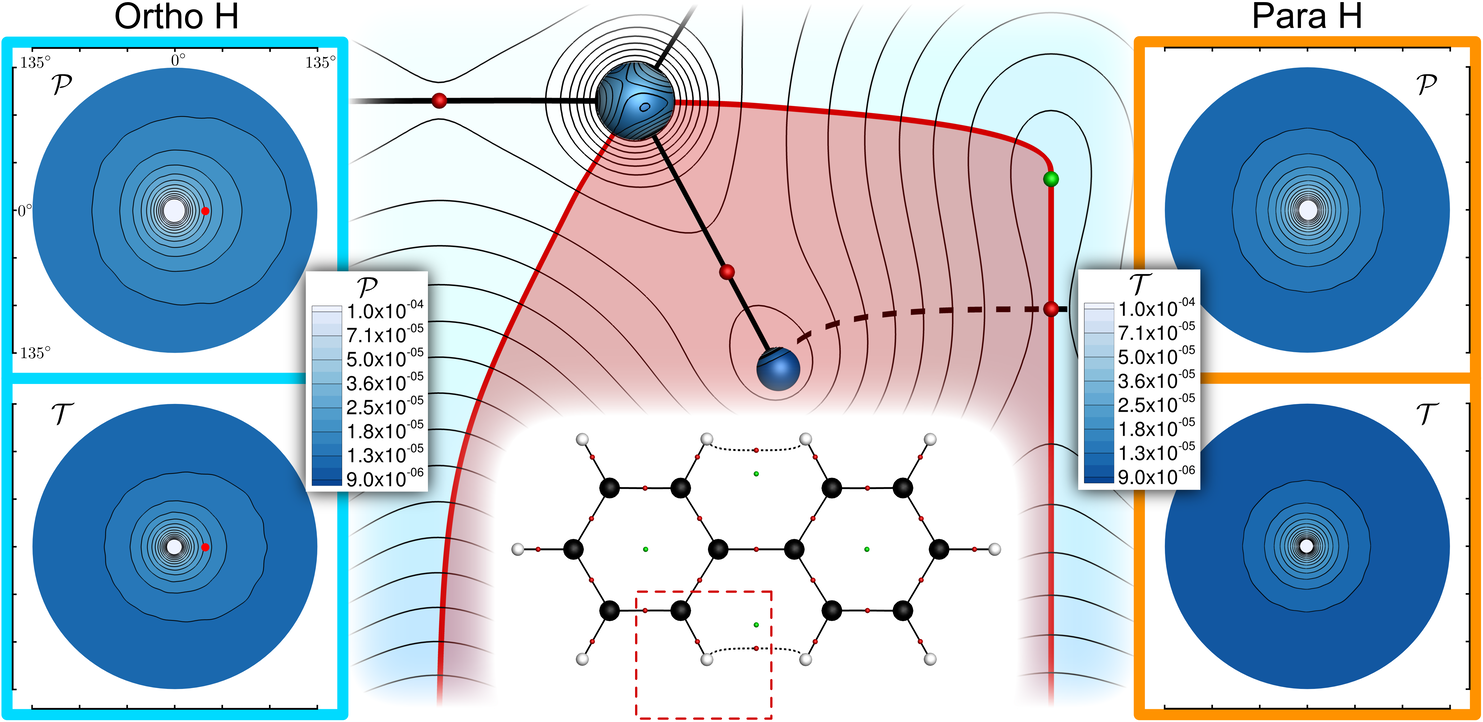}
	\caption{
		Planar biphenyl with stereographic projections of \ac{P} and \ac{T} for the ortho (cyan boxes; left) and para (orange boxes; right) hydrogen atoms.
		Center image shows contours of \ac{rho} on the molecular plane (corresponds to the region indicated by a dashed red box in the inset) with \ac{P} mapped onto spheres centered at the ortho carbon and hydrogen nuclei.
		The red region shows where the ortho \bs{C}{H} bond bundle intersects the molecular plane.
		Stereographic projections are centered at the \bs{C}{H} bond path with the molecular plane passing horizontally through the projections.
		The intersection of the ortho \bs{H}{H} bond path with the sphere is indicated by a red dot.
	}
	\label{fig:biphenyl}
\end{figure*}

Though maxima in \ac{P} typically map to bond paths, this is not always the case---a feature that allows one to ``see'' bond formation during chemical reactions and to assess the relative energies associated with such processes.
Consider for example the Lewis acid base reaction between borane and ammonia,
\begin{center}
	\ce{BH3 + $:$NH3 -> H3B-NH3}
\end{center}
\cref{fig:BH3NH3_isolated} depicts the bond bundles and bond wedges of \ce{NH3} and \ce{BH3} alongside their respective \acp{P}.  


Ammonia's nitrogen atom is distinguished by four maxima in \ac {P} and hence four bond wedges. 
Three of the these share bond wedges with those from hydrogen atoms, yielding three \bs{N}{H} bond bundles.
The bond wedge centered on what we identify as the nitrogen atom's south pole does not share an interatomic surface with another atom, as one would expect of a lone electron pair.
Borane, on the other hand, is characterized by three shared bond wedges with hydrogen atoms yielding three \bs{B}{H} bond bundles.
These bond wedges intersect along gradient paths that map to minima in \ac{P} located at the boron atomic poles.

The reaction between borane and ammonia molecules was simulated by aligning the south pole of the nitrogen atom with the north pole of the boron atom at an initial \bs{B}{N} distance of 3.0\AA.
The reaction was allowed to proceed to the equilibrium \bs{B}{N} distance of 1.65\AA, forming the \bs{B}{N} bond bundle depicted in \cref{fig:H3B-NH3_product}.

The evolution of the bond bundle and its associated condensed properties along the reaction profile are depicted in \cref{fig:BH3NH3_b}.
Inspection of this figure reveals that at 3.0\AA~separation, while there is a \ac{bcp} and a bond path connecting the boron and nitrogen \acp{ncp}, the two atoms do not contribute to a common bond bundle because the \bs{B}{H} bond wedges account for the entirety of \ac{P} on the boron.
The lone pair bond wedge on the nitrogen atom persists but there is no corresponding bond wedge on boron, which, at its pole, instead hosts a \ac{P} minimum---as well as minima for both \ac{V}, and \ac{T}.

Bond bundle formation begins at a \bs{B}{N} distance of approximately 2.70\AA~with the near simultaneous development of maxima in \ac{P}, \ac{V}, and \ac{T} at the boron pole.
The maximum in \ac{P} necessitates a bond wedge on the boron atom and an accompanying \bs{B}{N} bond bundle that continues to evolve over the course of the reaction to encompass a greater portion of the interatomic region between the boron and nitrogen atoms.

Across all calculations we find a correlation between \ac{P}, \ac{V}, and \ac{T}, hinting at an energy-charge density structure relationship we will discuss in more detail in \cref{sec:content}.
However, for now recall that when the forces acting on the molecules are small, \ie\ when far apart and at the equilibrium separation, the virial theorem asserts that $E(\Omega) = -T(\Omega)$.
Taking $\Omega$ to be the \bs{B}{N} bond bundle, in principle the energy of the region corresponding to \bs{B}{N} bond is given by the integral of \ac{T} over the appropriate bond wedges on the boron and nitrogen atoms.
Without evaluating the integral, however, it is clear from \cref{fig:BH3NH3_b} that the contribution to the total energy from the boron bond wedge is stabilizing, as one would expect.
Moreover, the simultaneous formation of basins in \ac{P} and \ac{T} emphasizes that a bond wedge is also an energetic stabilization basin, in agreement with the common conception of bonding as the concentration of charge that results in stabilization of the system.

\subsection{Spurious bond points and paths}
\label{sec:h-h_bonds}

\begin{figure}[h!]
	\centering
	\includegraphics[width=\linewidth]{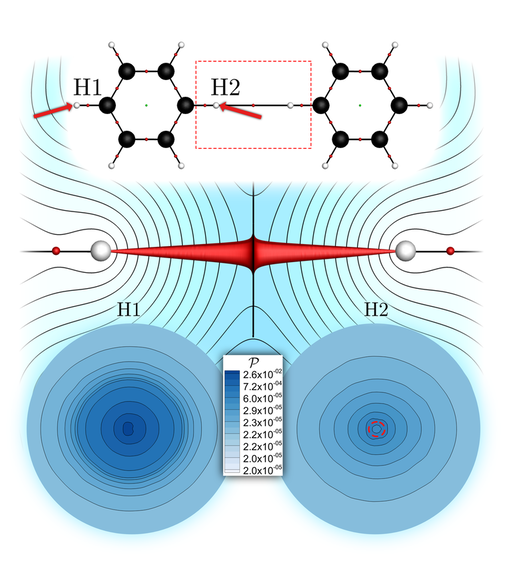}
	\caption{
		\bw{H}{H} bond bundle in planar dibenzene superimposed on contours of \ac{rho} in the molecular plane for the region indicated by the red dashed box in the inset.
		Stereographic projections of \ac{P} for two hydrogen atoms, centered opposite the \bs{C}{H} bond path (\ie\ for H2, at the \bw{H}{H} bond path; indicated by arrows in the inset).
		The bond wedge contributing to the \bw{H}{H} bond bundle is indicated by a slight maximum present at the center of the H2 projection (in a red dashed circle) that is absent in the H1 projection.
	}
	\label{fig:dibenzene}
\end{figure}

\begin{figure*}[h!]
\centering
\includegraphics[width=0.8\linewidth]{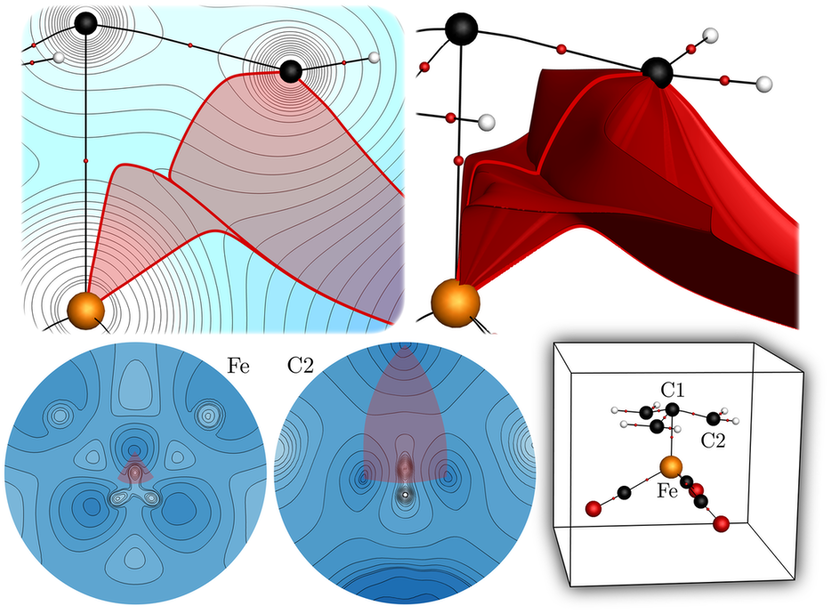}
\caption{
	Bond bundle in iron-trimethylenemethane for the \bs{Fe}{C2} interactions not accompanied by conventional \ac{qtaim} bond points and paths.
	\textbf{Top-left:} Contours of \ac{rho} on the Fe-C1-C2 cut plane, where the intersection of the bond bundle with the plane is shaded red, and the bond bundle surface intersections are indicated by red paths.
	\textbf{Top-right:} 3D rendering of the bond bundle with its intersection with the same cut plane indicated by a red path.
	\textbf{Bottom:} Stereographic projections of \ac{P} for Fe and C2, centered on the \bs{Fe}{C2} maxima and with the \bs{Fe}{C2} \ac{P}-basins shaded red.
}
\label{fig:iron-trimethylenemethane}
\end{figure*}

Another example in which the bond bundle picture provides more information than the conventional \ac{qtaim} representation is provided through an analysis of ``hydrogen-hydrogen'' bonding, as distinct from hydrogen or dihydrogen bonding \cite{matta2003}.
Two cases are considered here: i) the planar biphenyl molecule (\cref{fig:biphenyl}) characterized by \acp{bcp} between ortho hydrogen atoms that are not found in its twisted, lower energy conformation, and ii) dibenzene (\cref{fig:dibenzene}), also displaying unexpected \bw{H}{H} bond paths.
 
The first instance, planar biphenyl, has been thoroughly investigated. 
Some have argued that the \bw{H}{H} interaction is destabilizing and hence the occurrence of a bond path between the two hydrogen atoms is a failure of \ac{qtaim} \cite{Anti_qtaim2006b,Anti_qtaim2006a}. 
While others assert that the \bw{H}{H} bond path in fact lowers the energy of the metastable configuration \cite{matta2003,Matta2007_h-h,garcia-ramos2018,pendas2007}.

When analyzed from the bond bundle perspective, these arguments become moot.  
The \bw{H}{H} bond paths of biphenyl (\cref{fig:biphenyl}) do not map to maxima in \ac{P}.
These bond paths lie within the ortho \bs{C}{H} bond bundles.
Accordingly, there is no zero flux surface partitioning that exclusively constitutes the \bw{H}{H} bond path region.
Hence the energy of this region is ill-defined.

Second, the case of dibenzene in \cref{fig:dibenzene}, representative of the general complex of two molecules, \bw{\bs{R}{H}}{\bs{H}{R}}, between which \bw{H}{H} bond paths are found within some threshold intermolecular separation \cite{Anti_qtaim2006b}.
Here we do find a \bw{H}{H} bond bundle, allowing for a quantitative assessment of the \bw{H}{H} interaction.

The electron and kinetic energy counts of the \bw{H}{H} bond bundle are approximately $4\times10^{-3}$ electrons and 2 kcal/mol (less than $0.1~eV$), which amounts respectively to a 0.20\% and 0.26\% share of each hydrogen's 0.98 electrons and 380 kcal/mol of kinetic energy.
The energies of these bond bundles are on the same order as Van der Waals interactions. 

For such weak interactions, it is important to note the relationship between the analysis resolution, \ie\ the number of differential gradient bundles used, and the minimum solid angle that a \ac{P}-basin must occupy in order to be resolved.
For example, all the calculations for this work used (approximately) twenty-thousand \acp{dgb}, which means that a \ac{P}-basin whose solid angle is less than $1/20{,}000^{\rm{th}}$ of a steradian will not be recovered.
It is possible that \emph{every} bond path will map to a maximum in \ac{P} if a sufficiently high resolution is used.
Drawing from the \bw{H}{H} bond bundle of dibenzene (\cref{fig:dibenzene}), we expect the energy and electron counts of similar tiny bond bundles to be negligibly small.

If all bond paths \textit{do} map to maxima in \ac{P}, one could speculate that \acp{bcp} may function as nucleation sites about which bond bundles grow, not dissimilar to the nucleation sites in crystal growth.
Though this would change nothing about the instantaneous energetic significance of \acp{bcp} and bond paths associated with very small bond bundles, it would cast the \ac{bcp} itself as a spacial indicator of regions likely to become energetically significant.
The nature and extent of that significance could then be anticipated by the behavior of \ac{P} in the regions that correspond to bond paths.

\subsection{Missing bond points and paths}
\label{sec:missing_bonds}

Equally disconcerting to those seeking to rationalize molecular properties based on a \ac{qtaim} analysis are instances where \acp{bcp} are absent between atoms that, based on chemical reasoning, should be bound \cite{no_bcp_2006, no_bcp_2013a, no_bcp_2013b, no_bcp_2007, no_bcp_2009}. 

One such molecule is an iron trimethylenemethane complex represented in \cref{fig:iron-trimethylenemethane}. 
Based on simple bond counting schemes, interatomic distances, source function studies, delocalization indices, and spectroscopic evidence, the Fe atom and the methylene carbon atoms (C2 in \cref{fig:iron-trimethylenemethane}) should be bound \cite{no_bcp_2006}. 
Yet there is no \ac{bcp} between these atoms, and hence no bond path linking their \acp{ncp}.
As shown in \cref{fig:iron-trimethylenemethane} and consistent with the preponderance of the evidence and with chemical intuition, from a bond bundle perspective there is indeed a bond between these atoms.

\section{Condensed charge density and local kinetic energy}
\label{sec:content}

Though the local kinetic energy expressed in \cref{eq:Tgrad,eq:Tlap} are those most discussed in the literature \cite{Bader_KE, Bader1971, Ziff1977, Bader1978, Cohen1979, Bader1981b, Ghosh1984, Yang1996, Muga1997, Ayers2002b, Muga2005}, Anderson \ea\ \cite{Ayers_ambiguous_KE} have shown that these are but two representations drawn from an apparently infinite number of definitions for the local kinetic energy and correctly assert that ``{\it Regardless of how one partitions the system, it seems that the kinetic energy of an atom in a molecule is not uniquely defined in quantum mechanics: for any choice of subsystem partitioning, one can always find two 
$\dots$ functions that give different values for the regional kinetic energy.}'' 
While the authors go on to comment on the conceptual utility afforded through the use of \cref{eq:Tgrad,eq:Tlap}, or linear combinations of the two---constituting the Laplacian class of kinetic energy operators---they conclude, ``{\it [We] cannot think of any physical or intuitive justification for excluding local kinetic energies from outside the Laplacian family.}''

It is imperative to note that Anderson \ea\ were commenting on the quantum mechanical ambiguity associated with defining local kinetic energy and not reflecting directly on the value of approaches using one or another kinetic energy form.
Still, the implicit, and in some cases explicit \cite{Bader_on_chemistrry} expectation that QTAIM must be quantum mechanically rigorous has, in our opinion, compromised its utility.

Useful models often enlarge an existing conceptual framework, which for models of bonding means enhancing the chemist's view of charge density expansion and polarization.
These concepts are fundamental to many modern theories of bonding.
For example, {\sl ab initio} basis sets are specifically designed to capture the response of the charge density---and its underlying orbital basis---to expansion and polarization \cite{McQuarrie}.
Basis sets are even named in a manner that allows the user to quickly evaluate their ability to recover these charge density responses, \eg\ double zeta with polarization (DZP).
The degree to which \ac{rho} expands and polarizes upon bond formation is often used to categorize bonds as ionic, polar, covalent or polar covalent, and is frequently illustrated through the depiction of changes to charge density isosurfaces.
It is the central role of charge density isosurfaces in mediating the structure of \ac{P} that provides a rationale for representing the kinetic energy within the Laplacian family of energy operators.

Clearly, \ac{P} derives it structure most directly from the gradient field.
Consider an arbitrary gradient path \ac{gp} and a point $p$ on this path.
Through this point passes a charge density isosurface with its normal in the direction of the unit tangent vector to \ac{gp} at $p$, which we represent as $\bm{\tau}(p)$, \ie\,
\begin{equation}
\bm{\tau}(p) = \frac{\partial{\rm{G}}_i(\theta, \phi, s) }{\partial s} \bigg\rvert_{p}\\
\end{equation}

There are an infinite number of planes containing $\bm{\tau}(p)$.
Each of these {\it normal planes} intersects the isosurface along a plane curve, and in general each is characterized by a different curvature (not to be confused with the curvature of \ac{rho}).
The principal curvatures at $p$, denoted $\kappa_1(p)$ and $\kappa_2(p)$, are the maximum and minimum values of these curvatures.
The directions of these curves in the tangent plane at $p$ are referred to as principal directions, designated here as $\bm{e}_1(p)$ and $\bm{e}_2(p)$, with $\bm{e}_1(p) \cdot \bm{e}_2(p) = 0$.
An isosurface's lines of curvature are surface geodesics that are by definition always tangent to a principal direction.
There will be two orthogonal lines of curvature through every point on the surface that together act as a natural isosurface coordinate system.

Thus $\bm{\tau}$, $\bm{e}_1$ and $\bm{e}_2$, form an orthogonal moving frame ideally suited to describing charge density structure from a chemical perspective.
The fields $\bm{e}_1$ and $\bm{e}_2$ recover the shape of isosurfaces and $\bm{\tau}$ provides information as to the nesting of these surfaces.
 
\begin{figure}
	\centering
	\savebox{\tallerimage}{\includegraphics[width=0.49\linewidth]{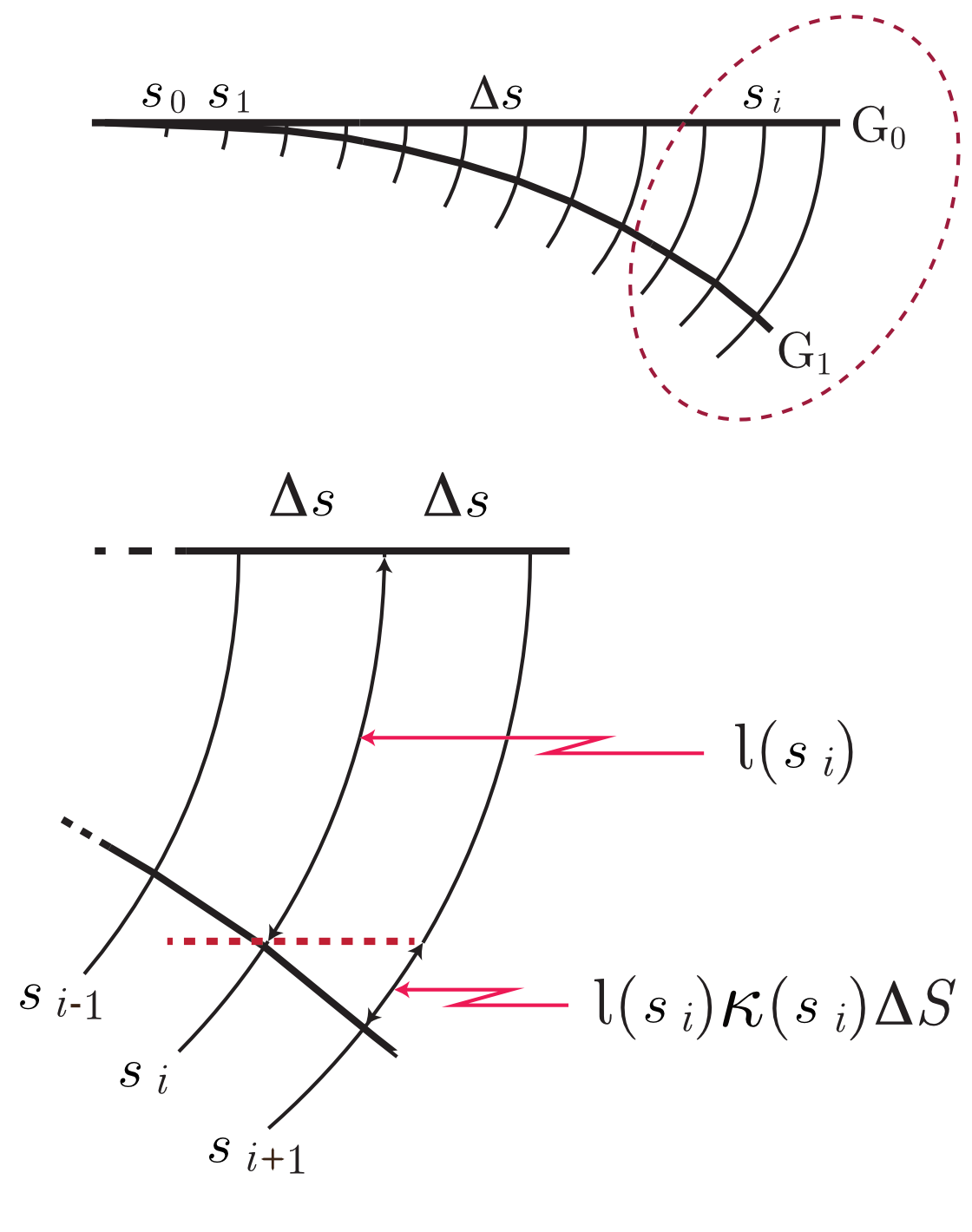}}%
	\subcaptionbox{\label{fig:GB_fig}}
	{\raisebox{\dimexpr.5\ht\tallerimage-.5\height}{\includegraphics[width=0.49\linewidth]{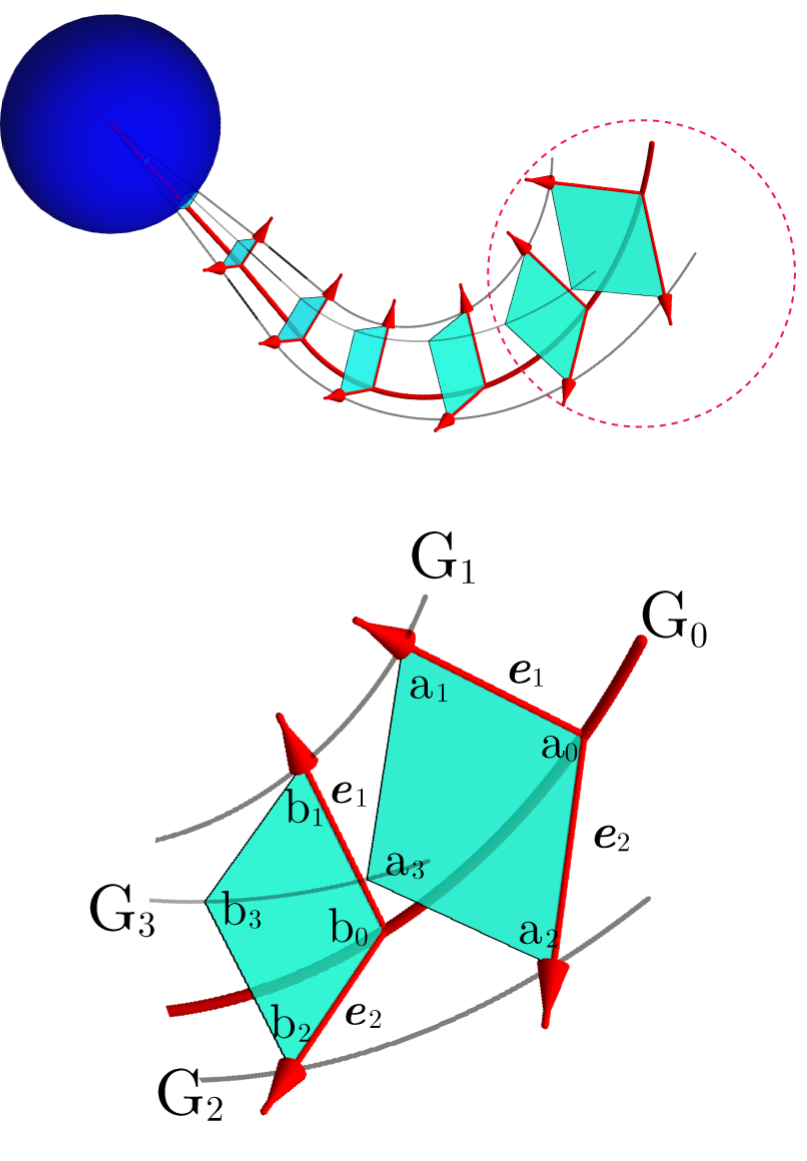}}}
	\subcaptionbox{\label{fig:ls}}
	{\usebox{\tallerimage}}
	\caption{%
		Diagrams illustrating the geometric behavior that determines the condensed charge density.
		\textbf{\subref*{fig:GB_fig})} Upper left: The set of tangent planes to isosurfaces about an atom in a molecule.
		The red arrows denote the directions of principal curvature at the points where $\rm{G_0}$ intersects the isosurfaces.
		\textbf{\subref*{fig:GB_fig})} Lower left: As per the text, the gradient paths $\rm{G_0}$, $\rm{G_1}$, $\rm{G_2}$ and $\rm{G_3}$ define the edges of a twisted rectangular pyramid (or in some cases a bipyramid) with faces formed from the union of the isosurfaces' principal curves passing through their point of intersection with $\rm{G_0}$.
		\textbf{\subref*{fig:ls})} Upper right: a 2D representation of one face of the rectangular pyramid showing principal curves on evenly spaced increments.
		\textbf{\subref*{fig:ls})} Lower right: The lengths of the principal curves in a face of the square pyramid obey a recursion relationship given by \cref{eq:recursion}.
	}
\end{figure}

Using the $(\bm{\tau}, \bm{e}_1, \bm{e}_2)$ moving frame, it is possible to derive the form of the differential area and volume elements introduced in \cref{eq:condensed_property}.
We begin by picking an arbitrary point $a_0$ on the arbitrary gradient path \ac{gp}$_0$ depicted in \cref{fig:GB_fig}.
(Here, consider the point to be located at roughly one third of the distance to the nearest atom.) This point lies on an isosurface ${\cal{S}}_a$.
Around this point construct a planar square patch by placing one of its vertices at $a_0$ and the remaining vertices at $a_1=a_0 + \epsilon \, \bm{e}_1(a_0) $, $a_2=a_0 + \epsilon \, \bm{e}_2(a_0)$ and $a_3=a_0 + \epsilon \, \big(\bm{e}_1(a_0) +\bm{e}_2(a_0)\big)$.
This patch is normal to $\bm{\tau}(a_0)$ and hence is a tangent plane to ${\cal{S}}_a$ at $a_0$.
For sufficiently small $\epsilon$ the edges of this patch are principal curves.
Passing through each vertex $a_i$ is a gradient path \ac{gp}$_i$.

Now pick another point on \ac{gp}$_0$, say $b_0$, which lies on the isosurface designated ${\cal{S}}_b$.
A vector originating at $b_0$ and lying in the direction $\bm{e}_1(b_0)$ will intersect \ac{gp}$_1$ at a point designated $b_1$ and in a similar fashion the vectors originating at $b_0$ and lying in the directions $\bm{e}_2(b_0)$ and $\bm{e}_1(b_0) + \bm{e}_2(b_0)$ will intersect \ac{gp}$_2$ and \ac{gp}$_3$ respectively at points designated $b_2$ and $b_3$.
The planar patch with vertices $b_i$ is necessarily a tangent plane to the isosurface ${\cal{S}}_b$ at $b_0$ with edges along principal directions. 

At every point $s_0$ along \ac{gp} there exists a coordinate patch (with vertices $s_i$) that is normal to $\bm {\tau} (s_0)$ and spanned by the vectors $\bm{e}_1(s_0)$ and $\bm{e}_2(s_0)$ such that \ac{gp}$_i$ is the union of all $s_i$.%
\footnote {This construction generates a set of triply orthogonal surfaces.
As with any such set, they necessarily intersect along lines of curvature.
}
Through every point in any of these patches there is a gradient path that maps one-to-one and onto each of these coordinate patches.
The differential gradient bundle may be taken as the union of all such coordinate patches. 

The charge contained in a differential gradient bundle can be found by integrating over the set of volumes formed by coordinate patches separated by $ds$, yielding a volume element equal to $ds\times dl_1\times dl_2$ where $dl_1$ and $dl_2$ are the lengths of the edges of the coordinate patch.

Obviously, $dl_1$ and $dl_2$ are functions of $s$, varying along the extent of the differential gradient bundle.
The form of this functionality may be discerned by considering a set of coordinate patches separated by a small distance $\Delta s$ (\cref{fig:ls}).
It is straightforward to show that for sufficiently small $\Delta s$,
\begin{equation}
\label{eq:recursion}
	l_1(s_{i+1} ) = l_1(s_{i} ) \big( 1 + \kappa_1(s_{i} ) \Delta s \big),
\end{equation}
where $\kappa_1(s_i)$ is the isosurface curvature in the $\bm{e}_1$ direction at $s_i$.
It follows that,
\begin{eqnarray}
	l_1(s_{n} ) =& \displaystyle{ \lim_{\Delta s\to0} l_1(s_{0} ) \prod_{i=0}^{n-1} \big(1 + \kappa_1(s_{i} ) \Delta s \big)}\\
	=& \hspace{-1.8em}\displaystyle{ l_1(s_{0} ) \exp \bigg( \int\limits_{s_0}^{s_n} \kappa_1(s) ds} \bigg).
\end{eqnarray}
Noting that $l_1(s_{0}) = d\theta \,s_0$ (see \cref{fig:dA}) we have,
\begin{equation}
	l_1(s)= \displaystyle{d\theta\, s_0\exp \bigg( \int\limits_{s_0}^{s} \kappa_1(s) ds} \bigg).
\end{equation}
Similarly, 
\begin{equation}
	l_2(s)= \displaystyle{d\phi \,s_0 \exp \bigg( \int\limits_{s_0}^{s} \kappa_2(s) ds} \bigg),
\end{equation}
and the area element, $dA(s) = l_1(s) \times l_2(s) $ is,
\begin{align}
	\label{eq:dA}
	dA(s) = & \quad \displaystyle{d\theta \,s_0 \exp \bigg( \int\limits_{s_0}^{s} \kappa_1(s) ds} \bigg)\displaystyle{d\phi \, s_0\, \exp \bigg( \int\limits_{s_0}^{s} \kappa_2(s) ds} \bigg) \nonumber \\
	= &\quad \displaystyle{d\theta \, d\phi \,s_0^2 \exp \bigg( \int\limits_{s_0}^{s} \big(\kappa_1(s)+\kappa_2(s) \big) ds} \bigg) \nonumber\\
	= & \quad \displaystyle{d\theta \, d\phi \, s_0^2 \exp \bigg( \int\limits_{s_0}^{s} 2H(s) ds} \bigg),
\end{align}
where $H(s)$ is the mean curvature, $ \big(\kappa_1(s)+\kappa_2(s) \big)/2$, of the isosurface at $s$.

Combining \cref{eq:condensed_property,eq:dA}, it is apparent that all gradient bundle condensed properties depend on the isosurface  mean curvature along \acs{gp}.
In particular, the gradient bundle condensed volume at a point $(\theta,\phi)$ of atom $i$ is,
\begin{equation}
	{\cal V}_i(\theta, \phi)=\displaystyle{\!\!\!\int\limits_{{\rm G}_i(\theta, \phi)} \!\!\! dA(s) ds}
\end{equation}
which is a characteristic of the charge density's extrinsic structure and may be thought of as measuring the total mean curvature of the isosurfaces along a gradient path.
The change in the magnitude of this quantity provides a direct measure of charge density expansion along a particular \ac{gp}, while its derivative properties with respect to $\theta$ and $\phi$ measure polarization.
 As an illustration consider the evolution of \ac{V} through the Lewis acid-base reaction discussed in \cref{sec:bh3nh3}.
 
\begin{figure*}
	\centering
	\includegraphics[width=.95\linewidth]{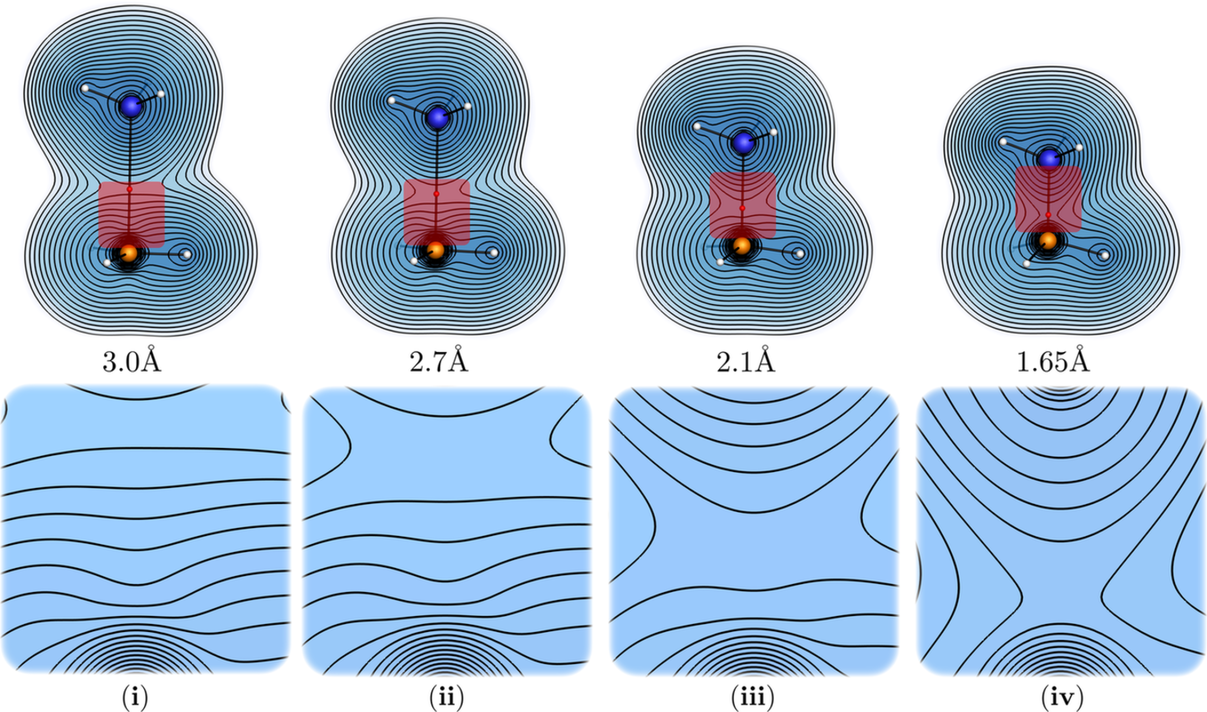}
	\caption{\label{fig:BH3NH3_contours}%
		The same contour levels along the \bs{B}{N} internuclear axis near the boron atom for the same four internuclear separations as in \cref{fig:BH3NH3_b}.
		Note the internuclear axis linking the boron and nitrogen atoms intersects contours of negative curvature in frames i-iii.
		Such regions cause neighboring \acp{gp} to converge and the natural volume element to diminish.
	}
\end{figure*}

\cref{fig:BH3NH3_contours} depicts the change in the charge density contours of the developing \bs{B}{N} bond bundle through the chemical reaction (recall \cref{fig:BH3NH3_b}).
Before the bond bundle forms, the bond path intersects isosurfaces about the boron atom at points of both positive and negative mean curvature.
In fact, the majority of boron isosurfaces are concave along the bond path.
The fraction of concave isosurfaces decreases along with the \bs{B}{N} internuclear distance.
At the equilibrium separation all boron isosurfaces are convex and a noticeable smooth corner has developed along the full length of the \bs{B}{N} bond path.

The behavior represented in \cref{fig:BH3NH3_contours} results from an increase in $p_z$ admixture from the boron atom.
Initially, the $p_z$ atomic orbital is essentially empty, becoming partially occupied through overlap with the lone pair orbital of the nitrogen atom.
Simply, the changing mean curvature along the bond path results from charge density expansion and polarization, and the total mean curvature as given by \ac{V} and \ac{P} serves as a measure of this response.

\section{Metallic bonds}
\label{sec:metal_bonds}

Real space models of metallic bonding date back to the earliest days of quantum mechanics.
Pauling proposed that interatomic forces and metallic structure could be rationalized from a resonating-valence-bond perspective {\cite{Pauling_1938, Pauling_1949}}.
Altmann \ea\ \cite{Altmann_1957} employed directed valence bond approaches in an attempt to explain the preferred crystal structure of the non-magnetic transition metals.
At nearly the same time, Engel and subsequently Brewer \cite{hume-rothery1968}, based largely on correlations, suggested that the spherically averaged number of valence $s$-$p$--electrons was the determining factor favoring one metal structure over another, an assertion deemed deficient by none other than Hume-Rothery {\cite{hume-rothery1965}.
Much more recently, Hoffmann \cite{Hoffmann_JT} unfolded band diagrams as a means of explaining the crystal structure preference of transition metals. 

Setting aside the highly focused work of Hoffmann, none of the other investigations produced insights as useful as the chemical models from which the various researchers took inspiration.
However, it is notable that in all cases the goal was to further chemical understanding of crystallographic structure.
Historically and practically, the ability of chemical models to account for molecular and solid state structure has served as an acid test for their continued exploration.
It is for this reason we compare the bond bundles of the BCC metal Nb with the FCC metal Cu as a way to demonstrate the potential of gradient bundle analysis.
A more complete comparison of bond bundles across the transition metals will be the subject of an upcoming paper. 

\begin{figure*}
	\centering
	\subcaptionbox{\label{fig:Cu_Nb_P}}
	{\includegraphics[width=0.49\linewidth]{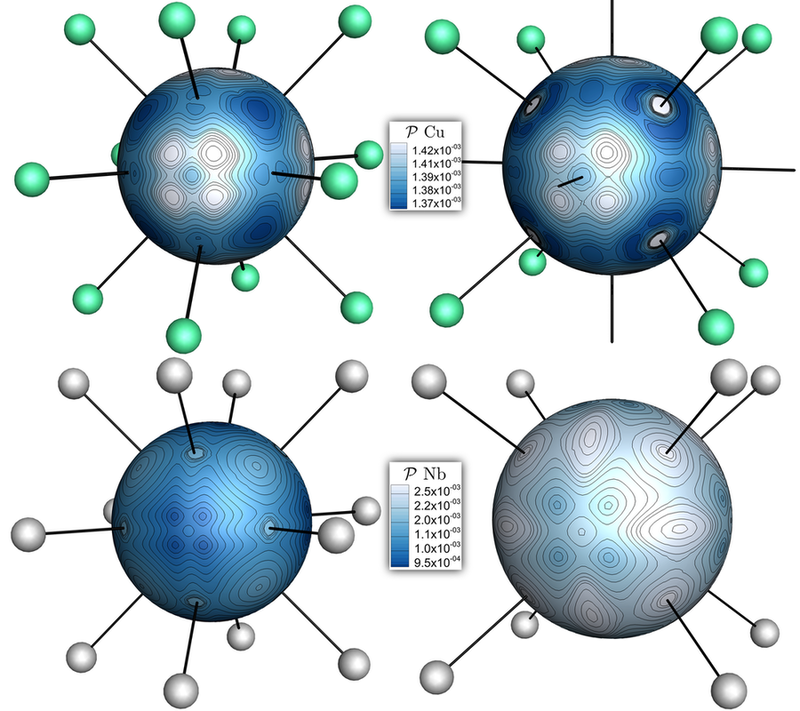}}
	\subcaptionbox{\label{fig:Cu_Nb_T}}
	{\includegraphics[width=0.49\linewidth]{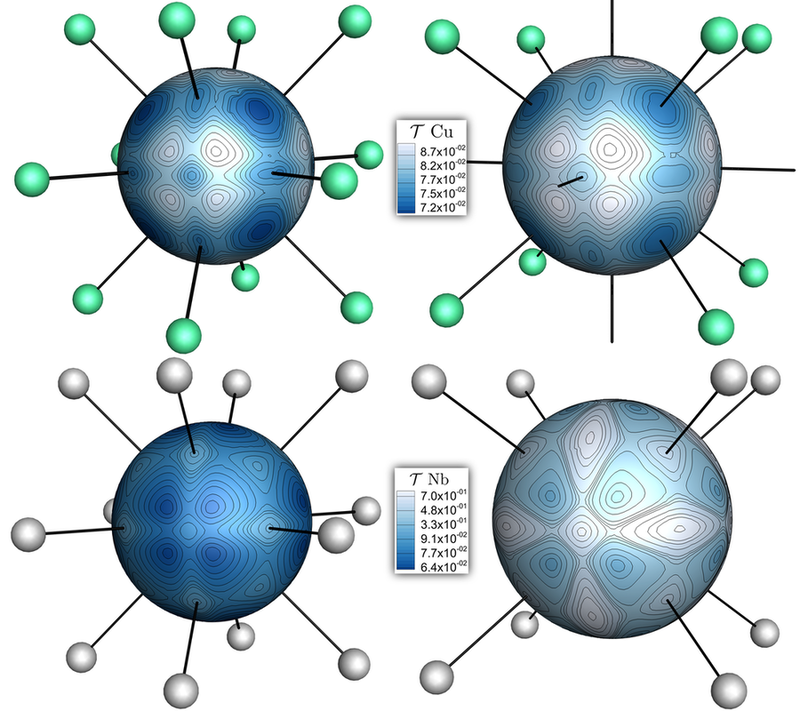}}
	\caption{
		\label{fig:Cu_Nb}%
		Condensed charge densities and condensed kinetic energy densities for FCC and BCC Cu and Nb.
		\textbf{\subref*{fig:Cu_Nb_P})} The condensed charge density, \ac{P} of FCC (left) and BCC (right) Cu and Nb, and 
		\textbf{\subref*{fig:Cu_Nb_T})} the condensed kinetic energy, \ac{T} of FCC (left) and BCC (right) Cu and Nb.
		The view is along the [100] direction looking from one atom through the octahedral hole toward its second neighbor.
	}
\end{figure*}

\cref{fig:Cu_Nb} depicts \ac{P} and \ac{T} for both Cu and Nb in FCC and BCC crystal structures.
Though the stable structure of Nb is BCC and that of Cu is FCC, both structures of each metal were modeled as a way of comparing stable and unstable gradient bundle condensed properties. 

As with the previous molecular calculations, there is a strong correlation between \ac{P} and \ac{T}.
(A weaker correlation with \ac{V} is not shown.) Remarkably, \ac{P} appears to reveal the underling $d$-orbital character of \ac{rho}, with Cu and Nb maxima capturing the nodal character of atomic orbitals that transform as the irreducible representations T$_{2g}$ and E$_g$ respectively. 
Though the magnitude of condensed property features for both \ac{P} and \ac{T} change dramatically with crystal structure, their topologies do not.
A fact that is readily explained by the observation regarding the orbital characteristics of gradient bundles.
On the one hand, the topology of \ac{P} is principally controlled by the distribution of electrons between E$_g$ and T$_{2g}$ atomic orbitals---a symmetry property (O$_H$ for both BCC and FCC).
On the other hand, the magnitude of \ac{T} and hence \ac{P} is controlled by the nature of the $\sigma$, $\pi$ and $\delta$ overlap between these atomic orbitals---a crystallographic property. 

With the virial theorem satisfied at every point of the condensed charge density, local maxima in kinetic energy correspond to regions of low total energy and hence are stabilizing relative to neighboring regions.
Accordingly, \cref{fig:Cu_Nb_T} reveals that the E$_g$-like gradient bundles stabilize the BCC structure of Nb, while T$_{2g}$-like gradient bundles stabilize FCC Cu. 

\begin{figure}
	\centering
	\includegraphics[width=\linewidth]{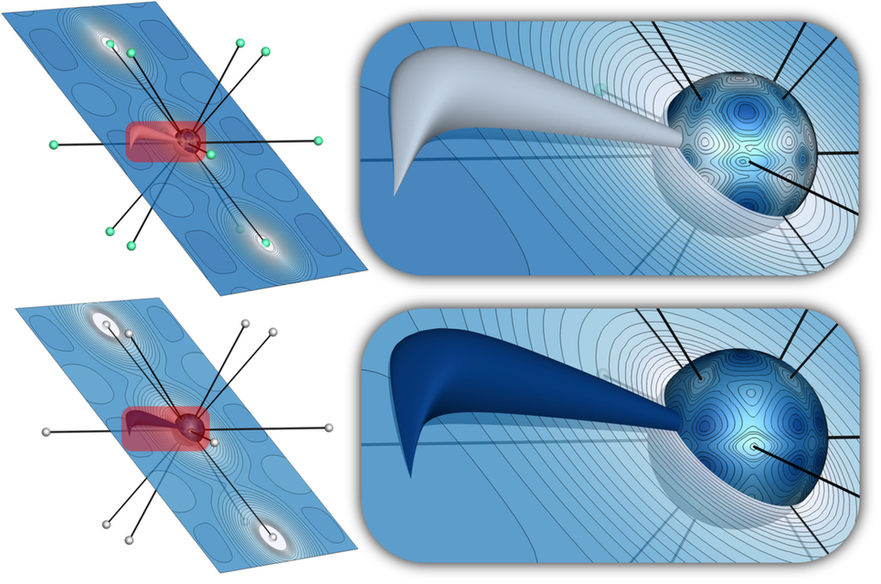}
	\caption{\label{fig:fcc_claw}%
		Gradient bundles shown with contours of \ac{rho} on the [110] plane.
		\textbf{Top:} The gradient bundle enclosing the high kinetic energy region of FCC Cu.
		\textbf{Bottom:} The gradient bundle encompassing the same region but for FCC Nb this is a low kinetic energy region.
	}
\end{figure}

One ``lobe'' of the high kinetic energy T$_{2g}$-like gradient bundles most responsible for stabilizing the FCC structure of Cu is represented in \cref{fig:fcc_claw} alongside the gradient bundles emanating from the same region of FCC Nb.
As expected, these gradient bundles are similar in shape.
However, for Nb, with five valence electrons, they emerge from local minima in \ac{P} and hence are ``under occupied'' compared to neighboring gradient bundles.
For Cu, with eleven valence electrons, they emerge from local maxima in \ac{P}, indicating that they are preferentially occupied by the valence electrons added when traversing the transition metal series from group 5 to group 11.
Of course, such preferential occupation is driven by a lower total energy---higher kinetic energy---with respect to all other gradient bundle occupations for the same number of electrons.
All in all, condensed charge density accumulates in the gradient bundles that will maximally lower the system total energy.
The correlation between \ac{P} and \ac{T} merely confirms the well-defined energy of gradient bundles. 

\begin{figure}
	\centering
	\includegraphics[width=\linewidth]{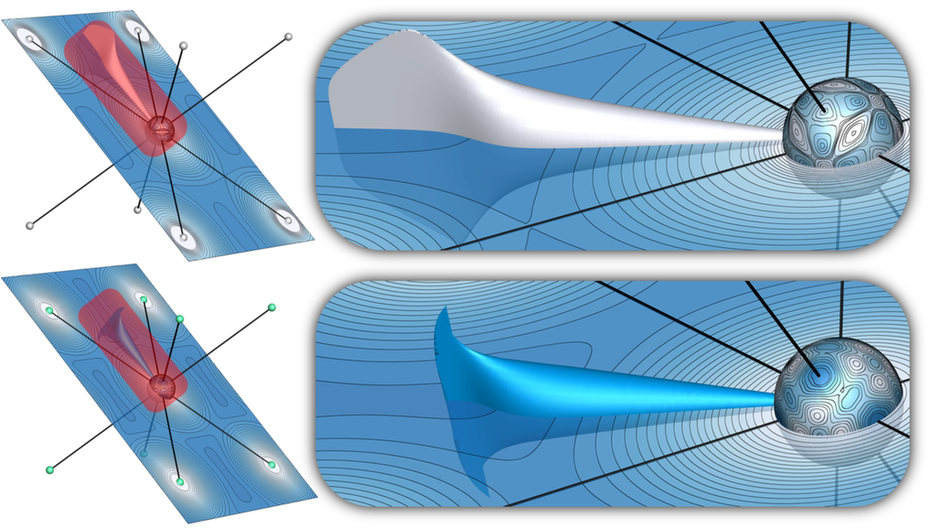}
	\caption{\label{fig:bcc_claw}%
		Gradient bundles shown with contours of \ac{rho} on the [110] plane.
		\textbf{Top:} The gradient bundle encompassing the high kinetic energy region of BCC Nb. 
		\textbf{Bottom:} The gradient bundle encompassing the same region, though for Cu this is a low kinetic energy region.
		Significantly, the gradient bundles have similar shape.  
	}
\end{figure}

In an analogous fashion, the high kinetic energy, E$_g$-like, gradient bundles most responsible for stabilizing BCC Nb are represented in \cref{fig:bcc_claw} alongside those emanating from the same region of BCC Cu.
Again these gradient bundles are similarly shaped though obviously they are preferentially filled by the first valence electrons of the transition metal series. 

We speculate that the preferential filling of gradient bundles is due to their relative content of bonding and antibonding character.
To illustrate, \cref{fig:bonding_antibonding_claws} shows early and late filling gradient bundles superimposed on representative bonding and antibonding molecular orbital contour diagrams.%
\footnote{The orbital contours are from large cluster calculations simulating the FCC and BCC structures.
It has been demonstrated that the central atom of such large clusters possesses an energy within a fraction of an eV of that from a bulk calculation and that the resulting charge densities are indistinguishable from those resulting from bulk calculations \cite{metal_molecules}.
}
The variation in the contour curvature is controlled by orbital nodal character.
The contour lines near nodes have a small mean curvature and experience their maximum mean curvature along gradient paths that are at greatest distance from nodes. 

\begin{figure*}
	\centering
	\subcaptionbox{\label{fig:Nb_bcc_clawMo}}
	{\includegraphics[width=0.48\linewidth]{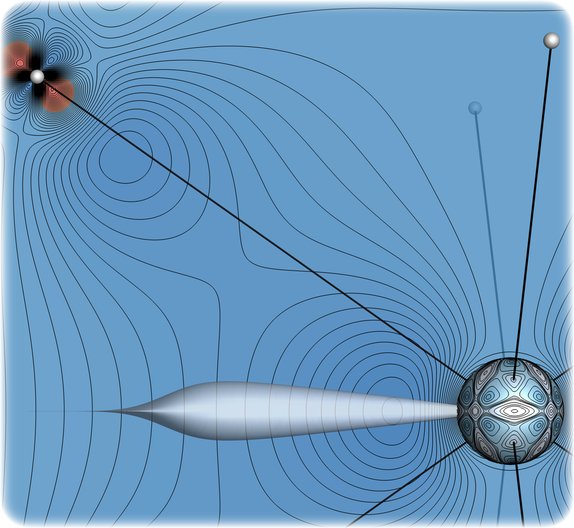}}\hspace{0.02\linewidth}
	\subcaptionbox{\label{fig:Cu_fcc_clawMO}}
	{\includegraphics[width=0.48\linewidth]{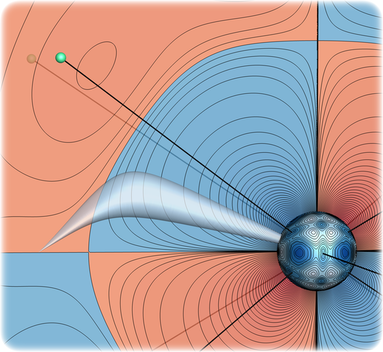}}
	\caption{
		\label{fig:bonding_antibonding_claws}%
		Illustrations of significant gradient bundles in FCC Cu and Nb and BCC Cu overlaid on contours of representative molecular orbitals on the [110] plane.
		Background colors indicate the relative phases of the molecular orbital wavefunctions.
		\textbf{\subref*{fig:Nb_bcc_clawMo})} The stabilizing gradient bundle of BCC Nb superimposed on a representative bonding orbital near the bottom of the $d$-band. 
		\textbf{\subref*{fig:Cu_fcc_clawMO})} The stabilizing gradient bundle of FCC Cu superimposed on a representative antibonding orbital near the top of the $d$-band.
			}
\end{figure*}

As a consequence, in the case of bonding orbitals it is the orbital gradient paths
(distinct from \ac{rho} gradient paths) more or less aligned with orbital antinodes that intersect the interatomic surface along orbital contours of maximum curvature (\cref{fig:Nb_bcc_clawMo}).
For antibonding orbitals the situation is quite different.
The orbital contours near the interatomic surface possess low mean curvature, as do those near the orbital's intrinsic angular nodes.
Such constraints force the formation of a curvature corner along the orbital gradient path roughly bisecting the interatomic and angular nodes (\cref{fig:Cu_fcc_clawMO}}). 

The gradient field of \ac{rho} must reflect the character of its orbital basis.
Gradient bundles containing predominately bonding character will approach the interatomic surface along nearly normal directions, while gradient bundles containing antibonding character will approach the surface with a tangential component.  

\begin{figure}
	\centering
	\includegraphics[width=0.98\linewidth]{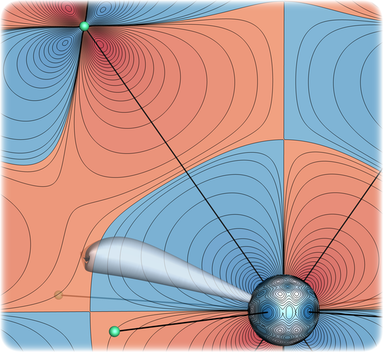}
	\caption{\label{fig:cu_partial_claw}%
		The stabilizing gradient bundle for BCC Cu superimposed on an antibonding orbital similar to that shown in \cref{fig:Cu_fcc_clawMO} on the [110] plane. Here, the molecular orbital contours along the  corner of intersection between the angular and interatomic  node are less curved than in  the  FCC case.  The reduced curvature is due to the weakened antibonding interaction around the  more open BCC octahedral hole.  With this interaction weakened the  stabilizing gradient bundle cannot harvest sufficient energy and does not curve  into the  the octahedral hole.
	}
\end{figure}

As illustrated in \cref{fig:bonding_antibonding_claws}, the gradient bundles stabilizing BCC Nb intersect regions that are predominately bonding and actually increase their volumes in these regions by intersecting orbital contours along paths near high mean curvature maxima.
The gradient bundles stabilizing FCC Cu maximize their volume in antibonding regions.
Leading to the seemingly paradoxical argument that FCC Cu is stabilized by antibonding interactions.

The paradox is easily resolved by noting that the virial theorem requires there be a corresponding decrease in the potential energy within the stabilizing gradient bundle.
This decrease results from a radial contraction of the charge density in the part of the gradient bundle closer to the nucleus, which screens that non-radial---curving---part of the bundle and allows it to expand into the high kinetic energy antibonding region.
Basically the gradient bundle of \cref{fig:Cu_fcc_clawMO} is harvesting excess kinetic energy from the antibonding region to stabilize the FCC structure. 

The previous conjecture is supported by considering the shape of the FCC stabilizing gradient bundle when Cu is forced BCC, as shown in \cref{fig:cu_partial_claw}.
Just as for the FCC structure, the gradient bundle projects into the crystallographic ``octahedral'' hole.
However, this interstice is less tightly packed when BCC, and in fact hosts a \ac{bcp} and bond path between second neighbors.
Obviously the larger expanse of the BCC versus FCC octahedral holes diminishes the intensity of antibonding interactions in the former, and accordingly the amount of kinetic energy that can be harvested from this hole.
The effect on the gradient bundle is evident, it does not curve and expand into the octahedral hole as it does in the FCC structure. 

We conclude that the stability of the FCC transition metals results from maximizing antibonding interactions, which is why the structure occurs late in the series where antibonding orbitals of the series are being filled.
The BCC structure is stabilized by gradient bundles that contain predominantly bonding character and therefore occurs early in the series.%
\footnote{
The reader may wonder why  the HCP structure occurs early (when predominately bonding orbitals are filled) and also just after the midpoint of the transition metal series when   antibonding orbitals are predominately filled. Though these two groups of transition metals have the same crystal structure, their charge density topologies are different and hence they possess different gradient fields and gradient bundle structures. \cite{jones2009a}
} 
The consequence of each structure is that its respective stabilizing gradient bundles contain maximal kinetic energy, thus maximizing the redistribution of charge and energy, yielding the minimum system total energy.

\section{Summary}
\label{sec:summary}

We have introduced a projected space that appears to be ideally suited to the investigation and discovery of charge density-property relationships in all classes of molecules and materials.
This space, denoted \ac{P}, constitutes all volumes bounded by zero flux surfaces in the gradient of the charge density.
As a consequence, when confined to the conventional Laplacian family of kinetic energy operators, each point of \ac{P} is endowed with a well-defined energy.
As a subset of the chemically meaningful structures of \ac{P}, all of those intrinsic to the \acl{qtaim} are recovered, \eg\ Bader atoms.
In addition, we also recover a class of nonstandard \ac{qtaim} structures that maximize charge density concentration and minimize energy, properties commonly associated with chemical bonds.  
These structures are called bond bundles.
We show that bond bundles are recoverable from \ac{rho} for any class of atomic system, provide more chemical information, and are not subject to the ambiguities and misgivings that have been associated with the \ac{qtaim} bond path.
We argue that the robust properties of bond bundles are rooted in the structure of \ac{P}, which is endowed with a natural reference frame that is determined by the system's charge density isosurfaces.
We demonstrate this fact by successfully analyzing transition metal structure within the bond bundle construct. 
 
This work should serve to further attempts at approaching traditional metallurgical problems from a chemical perspective. And also provides a new and possibly fruitful approach to analyze charge density in terms of its natural reference frame and the variations in this frame associated with chemical reactions and mechanical perturbations. 

\section*{Computational methods}
\label{sec:methods}

All simulations were performed with the \ac{adf} \cite{tevelde2001a,baerends1998, ADF2017authors} \textit{ab initio} software using the Perdew-Burke-Ernzerhof (PBE) functional \cite{perdew1996a} and a triple-zeta with polarization (TZP) all-electron basis set.
Calculation of \ac{P} was performed within the Tecplot 360 visualization package \cite{tecplotinc.2013} using the \acl{gba} tool of the in house Bondalyzer package by the \ac{mtg} at \ac{csm}.

\section*{Acknowledgments}
Support of this work under Office of Naval Research Grant No. N00014-10-1-0838 is gratefully acknowledged.
\bibliographystyle{unsrturl}
\bibliography{TheBond.bib}

\end{document}